\documentclass[12pt]{article}

\usepackage[sort]{natbib}
\usepackage{amsmath}
\usepackage{comment}
\usepackage{graphicx}
\usepackage{subcaption}
\usepackage[hmargin=2cm,vmargin={3.4cm,3.4cm},footskip=0.5in]{geometry}
\usepackage[usenames]{color}
\usepackage{url}
\usepackage[title]{appendix}

\usepackage[font=sf, labelfont={sf,bf}, margin=1cm]{caption}

\definecolor{Red}{rgb}{1,0.00,0.0}
\definecolor{Black}{rgb}{0,0.00,0.0}
\definecolor{purple}{rgb}{1,0.00,1}

\newcount\Comments  
\newcommand{\kibitz}[2]{\ifnum\Comments=1\textcolor{#1}{#2}\fi}

\allowdisplaybreaks

\begin{document}

\title{Modifying the Network-Based Stochastic SEIR Model to Account for Quarantine}

\author{Chris Groendyke \\
    Department of Mathematics\\
    Robert Morris University \\
    {\tt groendyke@rmu.edu}
  \and Adam Combs \\
    Department of Mathematics\\
    Robert Morris University \\
    {\tt combs@rmu.edu}
    }

\maketitle

\abstract{In this article, we present a modification to the network-based stochastic SEIR epidemic model which allows for modifications to the underlying contact network to account for the effects of quarantine.  We also discuss the changes needed to the model to incorporate situations where some proportion of the individuals who are infected remain asymptomatic throughout the course of the disease.  Using a generic network model where every potential contact exists with the same common probability, we conduct a simulation study in which we vary four key model parameters (transmission rate, probability of remaining asymptomatic, and the mean lengths of time spent in the Exposed and Infectious disease states) and examine the resulting impacts on various metrics of epidemic severity, including the effective reproduction number.  We find that the mean length of time spent in the Infectious state and the transmission rate are the most important model parameters, while the mean length of time spent in the Exposed state and the probability of remaining asymptomatic are less important. 
} \\


\normalsize

\newpage
\noindent
\section{Introduction}

In late 2019, the SARS-CoV-2, which is sometimes known colloquially as the novel coronavirus, and which is the virus which causes the COVID-19 illness, began to spread from Wuhan, China, to other cities, countries, and continents.  By mid-June 2020, it had reached almost every corner of the globe, infecting over 8 million people, and causing over 400,000 deaths \citep{nyt2020}, also resulting in wide-scale sociological and economic impacts.  Because a vaccine for this virus has yet to be developed, it is likely to continue to spread --- at least in some locations --- for at least several more months, and perhaps longer.  Thus, it is important to develop high-quality models to study the dynamics of the spread of this disease. 

Efforts to control and combat the spread of this epidemic have included strategies such as social distancing, self-isolation, and quarantine (among others), all of which are designed to alter the contact patterns of individuals in a population, and particularly those who are infected with the disease (even if they are unaware they are infected).  Due to these changes in contact patterns, particularly those only initiated by individuals upon developing symptoms of the disease, most of the models traditionally used to model the spread of epidemics, such as the standard SEIR (Susceptible-Exposed-Infectious-Removed) model are likely to be inadequate to model the progression of this epidemic.  It seems clear that modifications to the standard models are needed in order to sufficiently account for these new dynamics.

Because this novel coronavirus emerged only recently, the literature on modeling the spread of this virus is not yet robust or thorough.  However, several researchers have attempted to model the spread of this epidemic.  All or almost all of these efforts involve using compartmental disease models such as SEIR or modified versions thereof, along with a ``mean field'' model for the contacts between members of the population.  These mean field (sometimes called ``random mixing'' or ``well-mixed'') models, which have a long history in the study of disease dynamics  \citep{bailey1950sse, kermack1927cmt}, allow for any infectious member to infect any member of the susceptible class.  Under assumptions about the various disease parameters, it is often possible to express differential equations governing the sizes of the various compartments.  

Many researchers have taken just this approach to study the novel coronavirus.  \citet{iwata2020simulation} and \citet{wan2020will} conducted studies of this epidemic using standard random mixing SEIR models.  \citet{hou2020effectiveness} used a mean field SEIR model to describe this disease, and varied the rate of contacts to model the differences in transmission events caused by the effects of self-isolation.  \citet{peng2020epidemic} modified the mean field SEIR model by adding \textbf{Q} (Quarantined) and \textbf{P} (Insusceptible) classes.  \citet{lopez2020modified} took a similar approach by adding a \textbf{Q} class to represent the quarantining of infectious individuals, as well as a protected population compartment (\textbf{C}) to account for efforts to protect susceptible population members by confining them.  \citet{shi2020seir} added three new classes to the standard mean field SEIR model to account for the quarantining of individuals who are Susceptible, Exposed, and Infectious.  Similarly, \citet{wang2020evolving} added new classes to the mean field SEIR model to account for asymptomatic cases and the effects of hospitalization.  

There is, however, an alternative to the mean field approach of modeling interactions between population members which posits an underlying network describing the contact relationships between individuals.  In this framework, disease transmission can only occur between individuals who share an edge in this network.  This network approach can sometimes yield very different --- and in some cases more realistic --- dynamics for the spread of an epidemic \citep{meyers2005nta, keeling2005nae, ferrari2006network}.  Simulation studies using network models have been used in the past as an effective tool to investigate the properties of epidemics; this approach was taken in such works as  \citet{volz2008sdr} and \citet{barthelemy2005dpe}.  However, there has been little or no published work along these lines in the efforts to study the spread of the SARS-CoV-2 virus.  (We do note that \citet{prasse2020network} used a network model to study the spread of this disease; however, this contact network is between cities rather than individuals.)

Here we present a modification to the network-based SEIR model that has two key changes that make it more appropriate for the study of this disease.  First, there is an introduction of a separate class to model those individuals in self-isolation or quarantine; this is similar to some of the approaches above, but treats this new class as a subset of the infectious class.  (We will hereafter refer to this modified SEIR model as the SEI(Q)R model.)  Second, our network model allows for specific changes in network contacts upon the presence of symptoms (and also allows for a percentage of the infectious population to remain asymptomatic), which could be used to simulate the likely change in contact patterns of an individual in self-isolation or quarantine.  Thus, our network model is a dynamic, rather than static, network model, though the changes we allow in the network for this study are rather straightforward; see \citet{bansal2010dynamic} for a discussion of the role of dynamic contact networks in the study of disease dynamics.  In addition, we make use of stochastic models (as opposed to the deterministic ones sometimes utilized mean field models) to describe the lengths of time spent by individuals in the various disease stages, allowing for considerable flexibility in modeling, as well as a straightforward way to assess uncertainty.  We believe that this novel approach to studying SARS-CoV-2 provides a good framework for the investigation of this epidemic (particularly in smaller and/or closed populations) and is a viable alternative to the mean field models that have thus far been used to study the transmission of this disease.  

The remainder of this paper is organized as follows: in Section 2 we describe the epidemic and network models we use for this study, including the changes we make to the standard models to account for the effects of self-isolation and quarantine; in Section 3, we discuss how this model can be applied to study the spread of the SARS-CoV-2 virus, and conduct a simulation study to assess the sensitivity of the model to the various parameters; Section 4 concludes and offers ideas for future work and extensions of this model. 

\section{Epidemic and Network Models}
\label{sec:models}

\subsection{SEI(Q)R Epidemic Model}
\label{subsec:seiqr}

Compartmental models, i.e., models that assign each member of the population to one of a finite number of classes based on the state of the disease progression within that individual, have long been used to model the progress of various infectious diseases in individuals.  \citet{kermack1927cmt} initially introduced an SIR compartmental model.  This model consists of three classes: the Susceptible (\textbf{S}) class consists of those individuals who do not currently have the disease, but could contract it at some future point in time; the Infectious (\textbf{I}) class contains those individuals who can infect others in the population; the Removed (\textbf{R}) class consists of those members of the population who can no longer spread the disease to others.  Depending on the disease in question, it is sometimes assumed --- as will be the case here --- that once an individual enters the \textbf{R} class, they can never be reinfected, and hence play no further role in the progression of the epidemic through the population.

The SEIR model adds an Exposed (\textbf{E}) class, corresponding to individuals in the population who have contracted the disease, but cannot yet infect others.  Some early examples of this model include \citet{schwartz1983infinite}, \citet{aron1984seasonality}, and \citet{hethcote1980integral}.  This added class is a potentially important generalization of the SIR model, as it has been shown \citep{wearing2005appropriate} that failing to account for latent or incubation periods can lead to inaccurate or biased results.  The SEIR model has been used to model many types of infectious disease, including HIV/AIDS \citep{li2001global}, and measles \citep{momoh2013mathematical, grenfell1992chance}, as well as various types of influenza \citep{gonzalez2014fractional, dukic2012tracking, grais2003assessing}, and the novel coronavirus \citep{hou2020effectiveness, kuniya2020prediction}.  For a thorough review of compartmental epidemic models, as well as other approaches to modeling epidemics in populations, see \citet{keeling2011modeling}.

We propose a modification to the SEIR model in which a Quarantined (\textbf{Q}) class is added, as a subset of the \textbf{I} class.  This new class is designed to account for the fact that an individual becoming symptomatic with disease indicators --- which would happen at some point \textit{after} their becoming infectious --- may change their behavioral patterns in a way that impacts their interactions with others in the population.  That is to say, a person who becomes aware of their having disease symptoms may not behave similarly (e.g., they may self-isolate) to others in the \textbf{I} class who do not show symptoms and thus have no reason to believe they are infectious.  Thus, it seems reasonable to account for this sub-group of the \textbf{I} class separately; since the \textbf{Q} class is a sub-class of the \textbf{I} class, members of the \textbf{Q} class are still assumed to be able to infect others.  (Though they may have fewer opportunities to do so, since their number of contacts will typically be assumed to decrease; details of this are given in Section \ref{subsec:Q}.)  We assume that a person entering the \textbf{Q} class will remain there for the duration of the time they are infectious, and will then move to the \textbf{R} class, so that their progression would be \[S \rightarrow E \rightarrow I \rightarrow Q \rightarrow R\]

We also assume, however, that there is some proportion of the population who enter the \textbf{I} class who never enter the \textbf{Q} class; this feature is used to model the set of people who catch the disease, but never show any symptoms, and hence never have any reason to change their contact patterns.  (But it could also be used to model individuals who are infectious but choose for whatever reason not to alter their contact patterns.)  These individuals would then follow the standard progression through the SEIR model: \[S \rightarrow E \rightarrow I \rightarrow R\]

Finally, we note that in our model, the lengths of time spent in the \textbf{E}, \textbf{I}, and \textbf{Q} states are all stochastic, as is the time to infection along any given edge in the contact network.  The specifications of these random variables and accompanying parameters for this application are given in Section \ref{subsec:SEIQRcorona}.

\subsection{Network Structure}
\label{subsec:network}

The model we use to describe the structure of the contact network for the ($N$ individuals in the) population is a type of model known as a Bernoulli graph model whereby every pair of individuals in the population has a common probability $p$ of having a contact relationship.  This is a dyadic-independent model, meaning that the probability of an edge for a given dyad is independent of the presence or absence of edges among other dyads in the network.  This model was first described in \citet{gilbert1959random} and \citet{erdos1959rg}, and will be henceforth referred to as the Gilbert-Erd\H{o}s-R\'enyi (or GER) model.  This choice is intentionally simple and generic; it does not attempt to model any specific population.  It also minimizes network effects so that the effects of the disease model can more easily be examined.  Further, it can provide a baseline for comparison for future work that incorporates more complicated network structures.

An important aspect of our development is the recognition that the presence of the disease can change the nature of the contact network structure.  This is reflected in our model by specifying two different network parameters.  In particular, we will have one network parameter corresponding to the situation where neither member of the dyad is in the \textbf{Q} state (represented by $p$), and a second network parameter for dyads where one member of the dyad is the in \textbf{Q} state, and the other is not (represented by $p^\ast$).

Note that, for the purpose of simulating the spread of the epidemic through the population, it is not necessary to consider the case where both members of the dyad are in the \textbf{Q} state.  Recall that a transmission event can only occur from a member of the \textbf{I} class (which includes the \textbf{Q} class) to a member of the \textbf{S} class.  Thus, since any future disease transmission is impossible for two individuals who have entered the \textbf{Q} state, the presence or absence of an edge between these individuals cannot impact the progress of the epidemic.  We do note, however, that our method of estimating the reproduction number for an epidemic relies on the degree distribution of the network (see Section \ref{subsec:reproduction} for details).  For this reason, we do indeed consider a third network parameter, corresponding to \textbf{Q}-\textbf{Q} dyads, even though it has no impact on the actual spread of the disease in our model. 

\subsection{Modeling the Network Changes over the Course of the Epidemic} \label{subsec:Q}

At the point in the epidemic when an individual $i$ enters the \textbf{Q} state --- should such a transition occur --- we reconsider the network connections involving this individual.  Every dyad involving individual $i$, whether or not it currently has an edge in the contact network, is individually considered.  Dyads not including individual $i$ are ignored at this stage.  

At the time an individual $i$ enters quarantine, we will let $E$ denote the event that a given dyad (involving individual $i$) is connected by an edge, and $E^\prime$ be the complementary probability of this edge not existing.  We will also define $E^\ast$ as the event that such an edge exists after the network adjustments are made as a result of quarantine.  Let $p$ represent the \textit{a priori} probability of an edge between any individuals $i$ and $j$ when neither individual is in the \textbf{Q} state, and let $p^\ast$ represent the \textit{a priori} probability of such an edge when either $i$ or $j$ (but not both) is in the \textbf{Q} state.  

In order to get the desired expected post-quarantine edge density for individual $i$, we will condition on the pre-quarantine existence of an edge, giving
\begin{align}
p^\ast &= P[E^\ast]  \nonumber \\
	&= P[E^\ast | E] \cdot P[E] + P[E^\ast | E^\prime] \cdot P[E^\prime]  \nonumber \\
	&= P[E^\ast | E] \cdot p + P[E^\ast | E^\prime] \cdot (1-p) \label{eq:bayes1}
\end{align}

Because there is not a unique solution for the probabilities $P[E^\ast | E] \text{ and } P[E^\ast | E^\prime]$, we must impose a further condition.  There are many possible options, and the choice will depend on the nature of the phenomenon we are trying to model; for this application, we choose to set $P[E^\ast | E^\prime] = 0$.   This choice implies that no new edges are formed as a result of moving to the $\textbf{Q}$ state, and the only edges remaining for individual $i$ post-quarantine will be a subset of that individual's pre-quarantine edges.  Other choices are certainly possible; it might be reasonable in some cases to assume that there is some (presumably small) chance of forming new edges as a result of entering the \textbf{Q} state, though we disallow such possibilities here.

Then, under these assumptions, we can solve for the probability of a pre-quarantine edge involving individual $i$ remaining in the network as \begin{equation} P[E^\ast | E] = \frac{p^\ast}{p}\label{eq:edgeprob} \end{equation}

Note that in a typical application, we will have $p^\ast < p$, so that the ratio in Equation (\ref{eq:edgeprob}) will be less than 1.  However, this need not necessarily be the case, i.e., this model allows for situations in which $p^\ast > p$, and in this event, we could either bound this probability by 1 (and adjust the probability $P[E^\ast | E^\prime]$ accordingly to yield the correct edge density) or consider a different type of constraint on the conditional probabilities in Equation (\ref{eq:bayes1}) when making adjustments to the edges in the network.  The process for adjusting the network in the case where both individuals in a dyad are in the \textbf{Q} state is analogous.

\section{An Application to SARS-CoV-2}
\label{sec:app}

In this section, we describe how we apply the model discussed above to study the spread of the SARS-CoV-2 virus.  We first discuss the distributions and parameter values used to describe the lengths of time spent by individuals in the various states; we will refer to these as the \textbf{baseline} parameter values.  We then describe how to estimate reproduction numbers ($R_0$ and $R$) using our network and epidemic models.  Finally, through a simulation study, we analyze the impact of varying the different model parameters on statistics of interest with respect to the spread of the epidemic.

\subsection{Applying the SEI(Q)R Model to SARS-CoV-2}
\label{subsec:SEIQRcorona}

To determine our baseline model distributions and parameter values, we adapt the results of several previous studies; this is complicated somewhat by the fact that different researchers have modeled somewhat different quantities than the ones in our study.   \cite{He2020NatureMed} assumed the distribution for the \textit{incubation period} (length of time from exposure to becoming symptomatic) to be lognormal with mean of 5.2 days (see also \citealt{Li2020NEJM}).  This is similar to results in \cite{Linton2020JCM} which found the best fitting model for the incubation period to be lognormal with mean of 5.0 days.  \cite{He2020NatureMed} also estimated the number of days between the start of being infectious and the start of symptoms, that is the asymptomatic infectious period, to be about 2.3 days.   Combining these results, we use the lognormal distribution with mean of 2.9 days (and standard deviation of 2.51 days) to describe the of length of time from being Exposed to becoming Infectious, i.e., the length of time spent in the \textbf{E} state.  

It has been widely observed that some proportion of the individuals contracting this disease never develop symptoms.  For our baseline value of this asymptomatic percentage (which we label $\alpha$), we use the ``best estimate'' scenario given by the CDC \citep{cdc2020} of 35\%, which is similar to other estimates, such as \cite{nishiura2020estimation}.  For those individuals who do become symptomatic, we assume that the length of their asymptomatic infectious period, i.e., the length of time spent in the \textbf{I} state before entering the \textbf{Q} state, is uniformly distributed between 2 and 3 days, which is consistent with the mean estimate given by \cite{He2020NatureMed}.  The individuals who remain asymptomatic will skip the \textbf{Q} state and remain in the \textbf{I} state for the entirety of the time they are infectious.

For this study, we choose to use a gamma random variable with a mean of 12.5 days and a standard deviation of 5.0 days to describe the total length of time spent by an individual in the \textbf{I} state, to include any time spent in the \textbf{Q} state.  Estimates of the length of the infectious period for this disease have varied considerably, with most studies to date using the positive detection of a virus in an individual (for example, in the throat or stool) as a proxy for the individual being infectious.  For example, \citet{ling2020persistence} finds that the virus is detectable in throat samples for a median time of 9.5 days after symptom onset (so perhaps roughly 11 - 14 days after the individual has become infectious).

We use an exponential random variable to model the length of time taken for a transmission event across a given edge in the contact network, from an infectious individual to a susceptible one.  The reciprocal of the mean of this random variable is sometimes referred to as the ``transmission rate'' and is represented by $\beta$ in many models.  Because we model transmissions as occurring across a contact network (which is a substantially different model than has been commonly used to date for this virus) comparison of this model parameter is difficult.  However, two studies \citep{radulescu2020management, fang2020transmission} which have comparable interpretations of this parameter to ours both use a value of $\beta = 0.1$, which we also take as our baseline value for this parameter.

Finally, we set the values of the network parameters to $p = 0.047$ and $p^\ast = 0.011$ to account for the presumed tendency of individuals to reduce the number of people they are in contact with upon entering the quarantined state.  We set the probability of two individuals who are both in the \textbf{Q} state sharing an edge in the contact network at 0.0067; though as discussed above, this value does not impact the actual spread of the disease and is only needed for the calculation of the reproduction number at the end of the outbreak.  These values are somewhat arbitrary, as we are not trying to model the contact patterns of any specific population. 

\subsection{Estimation of Reproduction Numbers}
\label{subsec:reproduction}

When assessing disease dynamics, one of the metrics often used to describe the capacity of an epidemic to spread through a population is the \textbf{basic reproduction number}, $R_0$.  This is commonly defined as the number of disease transmissions that can be expected to be caused by an initial infected person in a population of otherwise susceptible individuals \citep{anderson1992infectious}.  As the disease spreads through the population, the number of transmissions caused by each infectious individual will tend to decrease, due to both the depletion of susceptible individuals (who have subsequently been infected), as well as any changes in the behaviors of individuals in response to the epidemic.  To account for this change through time, we also consider the \textbf{effective reproduction number at time $t$}, $R_t$, which is the expected number of secondary cases for an infectious individual at time $t$. 

For the network and epidemic models used here, \citet{groendyke2012network} developed the following formula that can be used to estimate $R$, based on previous work by \citet{andersson1998limit}, \citet{meyers2007contact}, and \citet{kenah2011contact}:

\begin{equation}
R = \left(\frac{E\left[D^2\right]}{E\left[D\right]} - 1\right) \cdot \left(1 - \left[\frac{1}{1+\beta\theta_I} \right]^{k_I} \right),
\label{eq:Rcalc}
\end{equation}
where $D$ is the random variable governing the degree distribution of the contact network, and $k_I$ and $\theta_I$ are the parameters of the gamma variable describing the length of time an individual remains in the \textbf{I} state.  Here, the change in the network structure due to quarantine will impact the degree distribution of the network, allowing us to assess the change in effective reproduction number from the start of the epidemic ($R_0$) to the end ($R_\omega$). 

\subsection{Simulation Study}
\label{subsec:simstudy}

This subsection gives the results of our simulation study.  We first examine the simulated epidemics produced at the baseline parameter values.  Then we analyze the impact of varying the transmission rate, probability of remaining asymptomatic, and mean lengths of time spent in the Exposed and Infectious states.  We vary each of these variables from 50\% to 150\% of its baseline value to encompass a range of reasonable values for each parameter, holding the values of all other variables constant.
 
\subsubsection{Simulations at Baseline Parameter Values}
\label{subsubsec:baseline}

\begin{figure}[h!t]
\begin{subfigure}{0.5\textwidth}
  \centering
\includegraphics[scale=0.45]{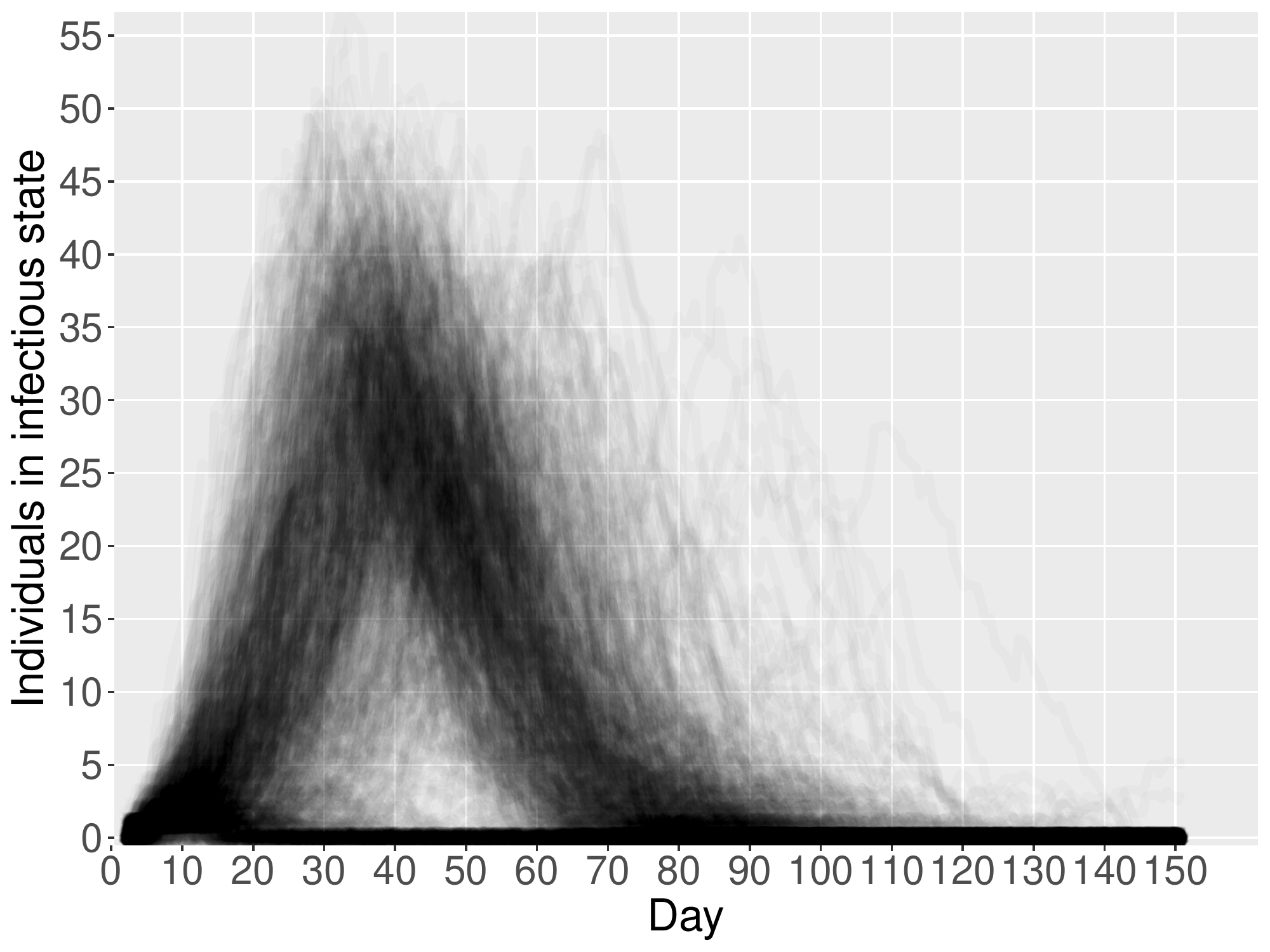}
\caption{Size of Infectious Class Over Time}  \label{fig:ninfsbase}
\end{subfigure}
\begin{subfigure}{0.5\textwidth}
\includegraphics[scale=0.45]{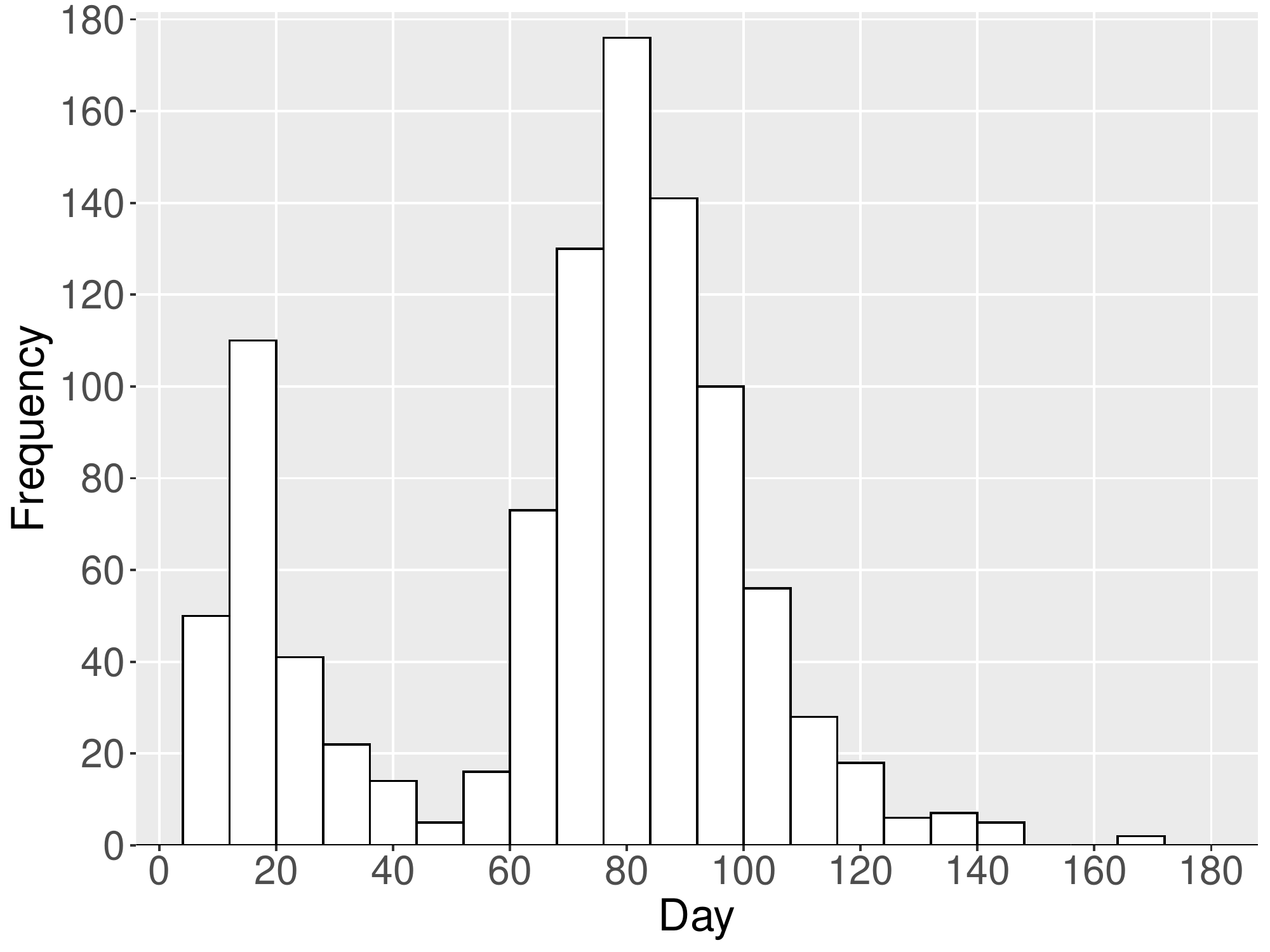}
\caption{Total Epidemic Lengths}  \label{fig:Rmaxbase}
\end{subfigure}
\caption{Trajectories of simulated epidemics under baseline parameter values.  Each light grey line in (a) represents the size of the \textbf{I} class over time for a single simulated epidemic.  A histogram of the total lengths of the epidemics is given in (b). }
\label{fig:baseplots}
\end{figure}

To establish a basis for comparison, we first run 1,000 simulated epidemics through a population of 100 individuals at the baseline parameter values.  To get a sense of the trajectory of the epidemic, consider Figure \ref{fig:ninfsbase}, which shows the number of individuals in the Infectious (\textbf{I}) class over time, for each of the simulated epidemics.  Most of the simulated epidemics follow a consistent pattern, with a relatively symmetric pattern of increase and decrease in Infectious class size, peaking around day 40.  (Time 0 is defined as the exposure time for the initially infected individual.)  The transmission tree for one sample simulated epidemic is given in Figure \ref{fig:transtree} in Appendix A.  The total lengths of the simulated epidemics are displayed in Figure \ref{fig:Rmaxbase}.  We note that about 15\% of the epidemics die out after the initially infected individual failed to infect any others, while another 6\% only infect one other individual; these cases largely account for the leftmost mode.  Of the other epidemics that spread to multiple individuals, the majority last from 70 - 90 days.

In many places, one of the biggest concerns of the COVID-19 pandemic has been the stress that it has placed on the local health care systems, with ICU beds, hospital staff, and ventilators being in short supply at various times.  In this study, we consider two different metrics as proxies for the level of stress induced on the health care system by the epidemic: the maximum size of the Infectious class (\textbf{I}$_{max}$), and the number of days the size of the Infectious class exceeds 15\% of the population (this threshold is arbitrary, but serves as a useful benchmark); we refer to the latter metric as ``stress days'' ($D_s$).

\begin{figure}[h!t]
\begin{subfigure}{0.5\textwidth}
  \centering
  \includegraphics[scale=0.45]{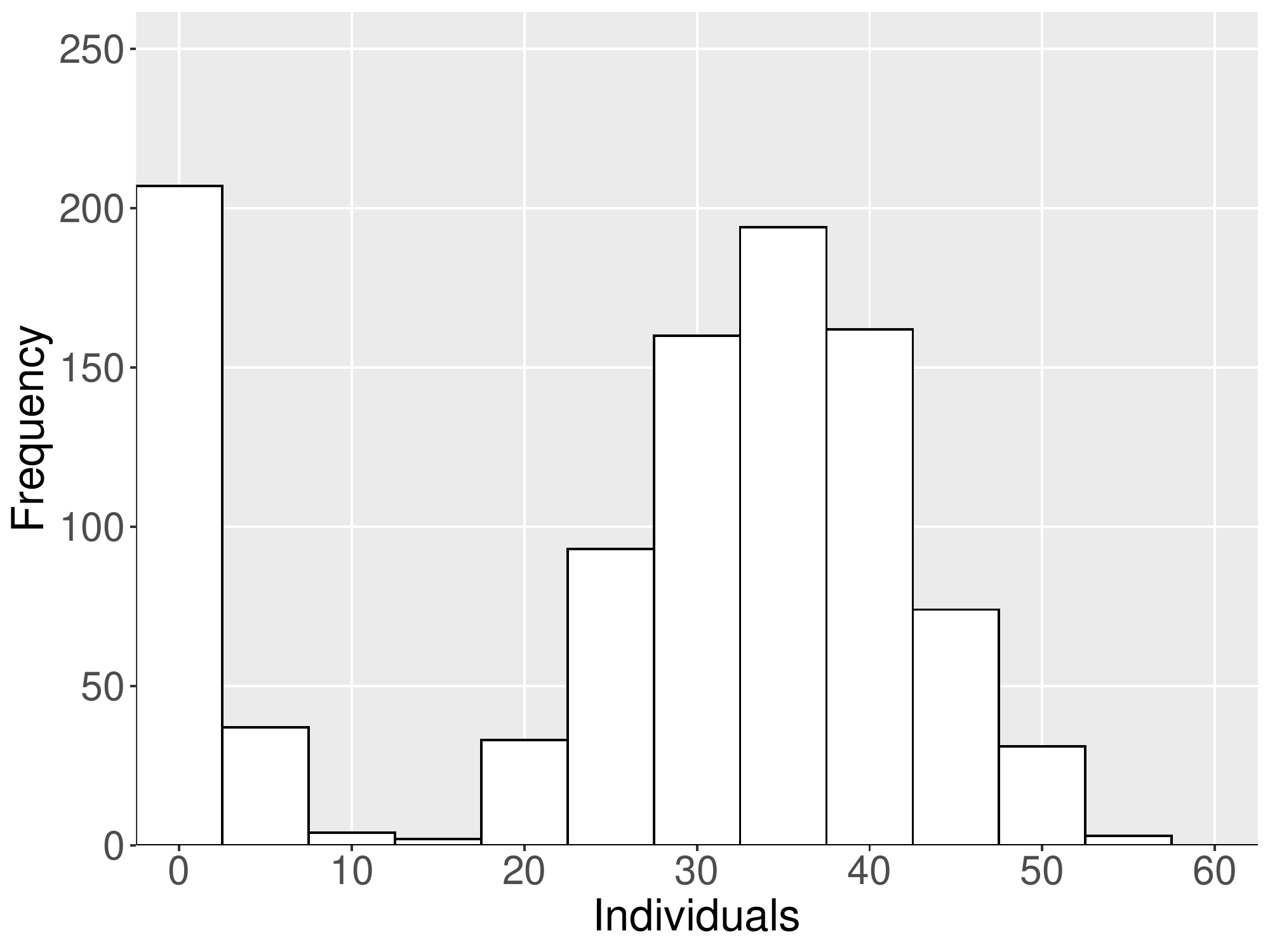}  
  \caption{Maximum Size of Infectious Class}
  \label{fig:sub-first}
\end{subfigure}
\begin{subfigure}{0.5\textwidth}
  \centering
  \includegraphics[scale=0.45]{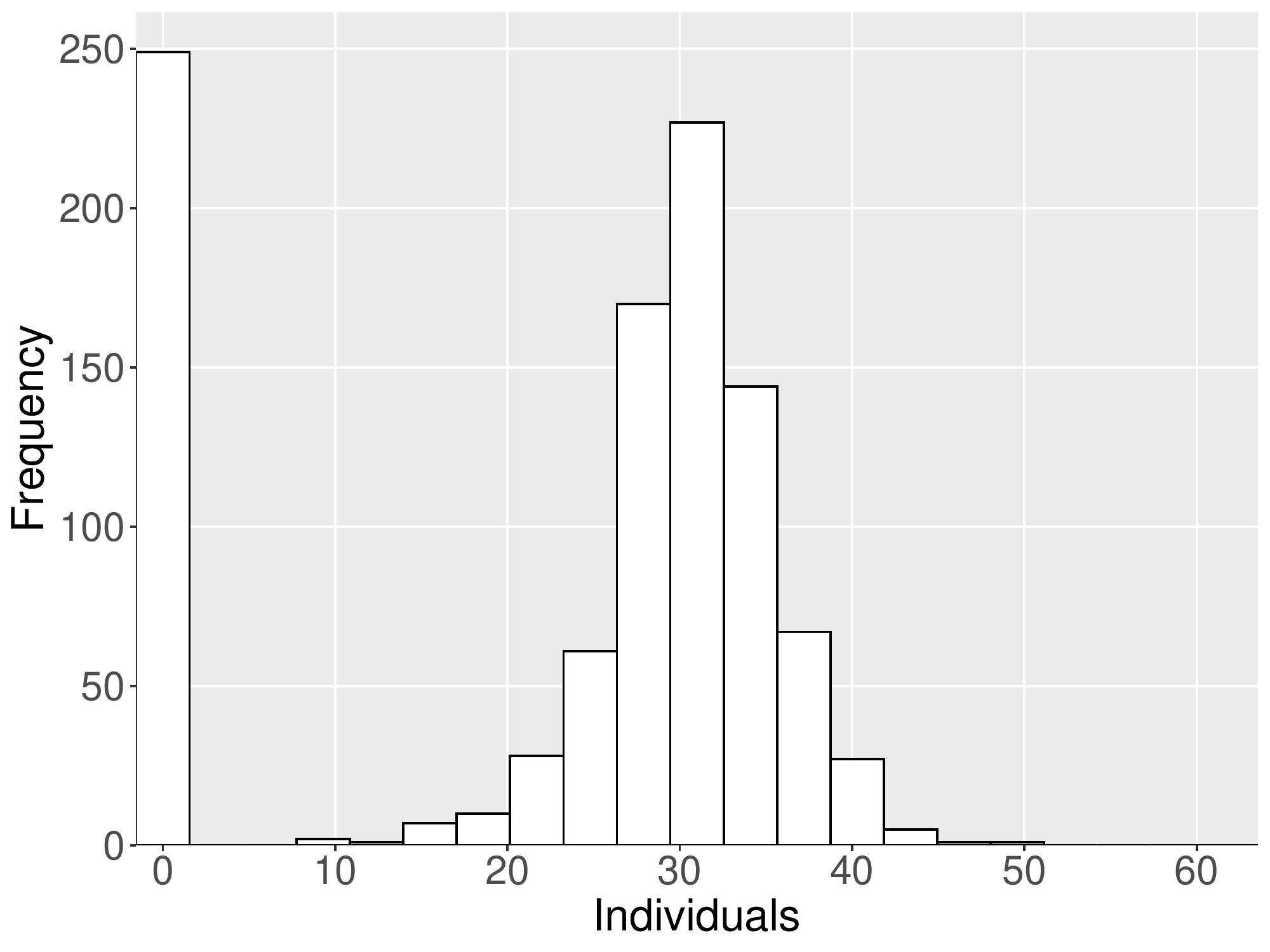}  
  \caption{Number of Stress Days}
  \label{fig:sub-second}
\end{subfigure}
\caption{Impact of epidemic on health care system under baseline parameter values.  Histograms of the maximum size of the infectious class (a) and the number of stress days induced by the epidemic (b).}
\label{fig:stressbase}
\end{figure}

Figure \ref{fig:stressbase} gives histograms of both of these metrics for the simulated epidemics at the baseline parameter values.  The size of the Infectious class tends to peak at around 25\% - 45\% of the population in most cases; we see that the Infectious class exceeds 15\% of the population for 25-40 days in the bulk of the simulated epidemics.

\begin{figure}[h!t]
  \centering
  \includegraphics[scale=0.45]{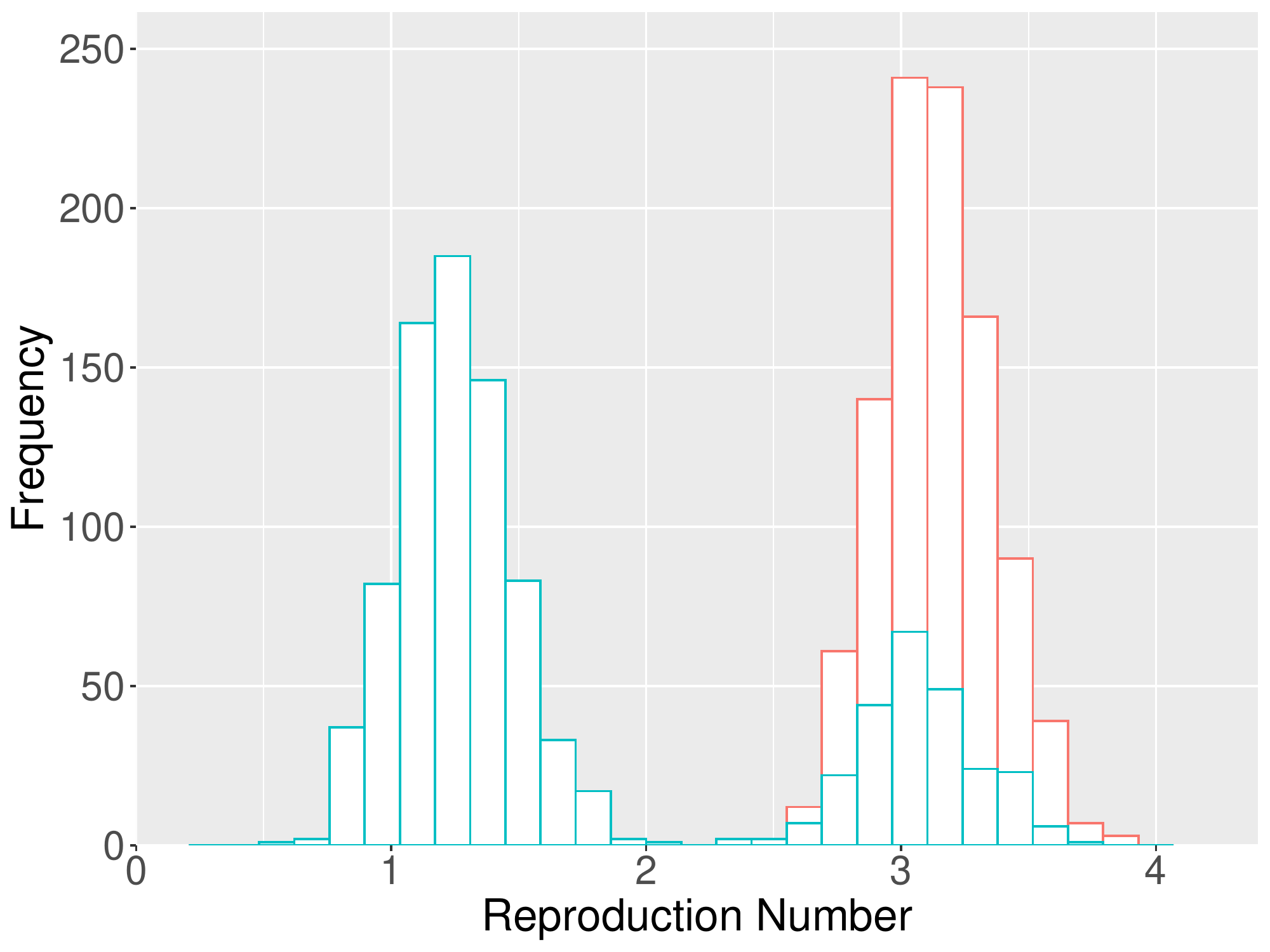}  
\caption{Histogram of estimated reproduction numbers for epidemics simulated under baseline parameter values.  Red represents the reproduction number at the start of an epidemic ($R_0$), while blue represents the reproduction number at the end of the epidemic ($R_\omega$).}
\label{fig:Rbase}
\end{figure}

We also calculate estimates of the effective reproduction numbers at the beginning and end of each of the simulated epidemics.  Figure \ref{fig:Rbase} gives a histogram of the calculated reproduction numbers pre- and post-epidemic for each of the simulated epidemics.  We can see that the reproduction number tends to decrease significantly from the start to the end of the epidemic, due to the effects of the quarantine.  Recall from Equation (\ref{eq:Rcalc}) that in our model, $R$ is a function of the degree distribution of the underlying contact network; as individuals lose contacts due to entering quarantine, their expected number of contact decreases significantly, hence bringing down the estimated value of the reproduction number.  We again see the effect of the portion of epidemics where only a single individual is infected; these are the cases where $R$ does not decrease significantly over the course of the epidemic.  We note that our estimated values of $R_0$, while broadly reasonable, are slightly higher than those found by some other researchers: \citet{Li2020NEJM} estimated a value of 2.2 based on the first 425 cases in Wuhan, China; \citet{nkwayep2020short} calculated a value of approximately 2.95 using data from Cameroon; \citet{liu2020reproductive} computed a median value of 2.79 from their meta-analysis of the reproduction number for this disease.  The bulk of our $R_0$ values can be seen to fall in the range of 2.8 - 3.7.  However, as mentioned above, in our model this calculation is impacted by the underlying contact network in the population, which we have chosen arbitrarily.  Thus, we do not expect the corresponding $R_0$ values to be realistic estimates for any particular population.  Rather, our main interests in calculating these values are to assess their change over the course of the epidemic, as well as to gauge how they vary with the model parameters.

\subsubsection{Varying the Transmission Rate}
\label{subsubsec:trans}

We first vary the transmission rate $\beta$, which describes how quickly the disease can be expected to spread across a given edge in the contact network, from an infectious individual to a susceptible one, from 0.05 to 0.15.  The lengths of the epidemics form a bimodal distribution for all values of $\beta$, similar to the baseline case (see Figure \ref{fig:Rmaxbase}).  For the smaller values of the transmission rate, the leftmost mode is more pronounced, while the opposite is the case for the larger values of $\beta.$  Table \ref{tab:Rmaxbeta} gives the mean and median epidemic lengths for the epidemics simulated under the various values of the transmission rate.

\begin{table}[h!t]
\centering
\begin{tabular}{lrrrrrrrrrrr}
  \hline
Transmission Rate & 0.05 & 0.06 & 0.07 & 0.08 & 0.09 & \textbf{0.10} & 0.11 & 0.12 & 0.13 & 0.14 & 0.15 \\ 
  \hline
Length (mean) & 53.0 & 63.7 & 69.2 & 71.7 & 71.9 & \textbf{70.0} & 69.2 & 67.7 & 67.4 & 65.8 & 64.5 \\ 
Length (median) & 34.0 & 54.5 & 80.5 & 82.0 & 81.0 & \textbf{78.0} & 75.0 & 73.0 & 71.0 & 69.0 & 67.0 \\ 
   \hline
\end{tabular}
\caption{Mean and median lengths (in days) of simulated epidemics for various transmission rates.  Values for the baseline parameter case are bolded.\label{tab:Rmaxbeta}}
\end{table}

\begin{figure}[h!t]
\begin{subfigure}{0.5\textwidth}
  \centering
  \includegraphics[scale=0.45]{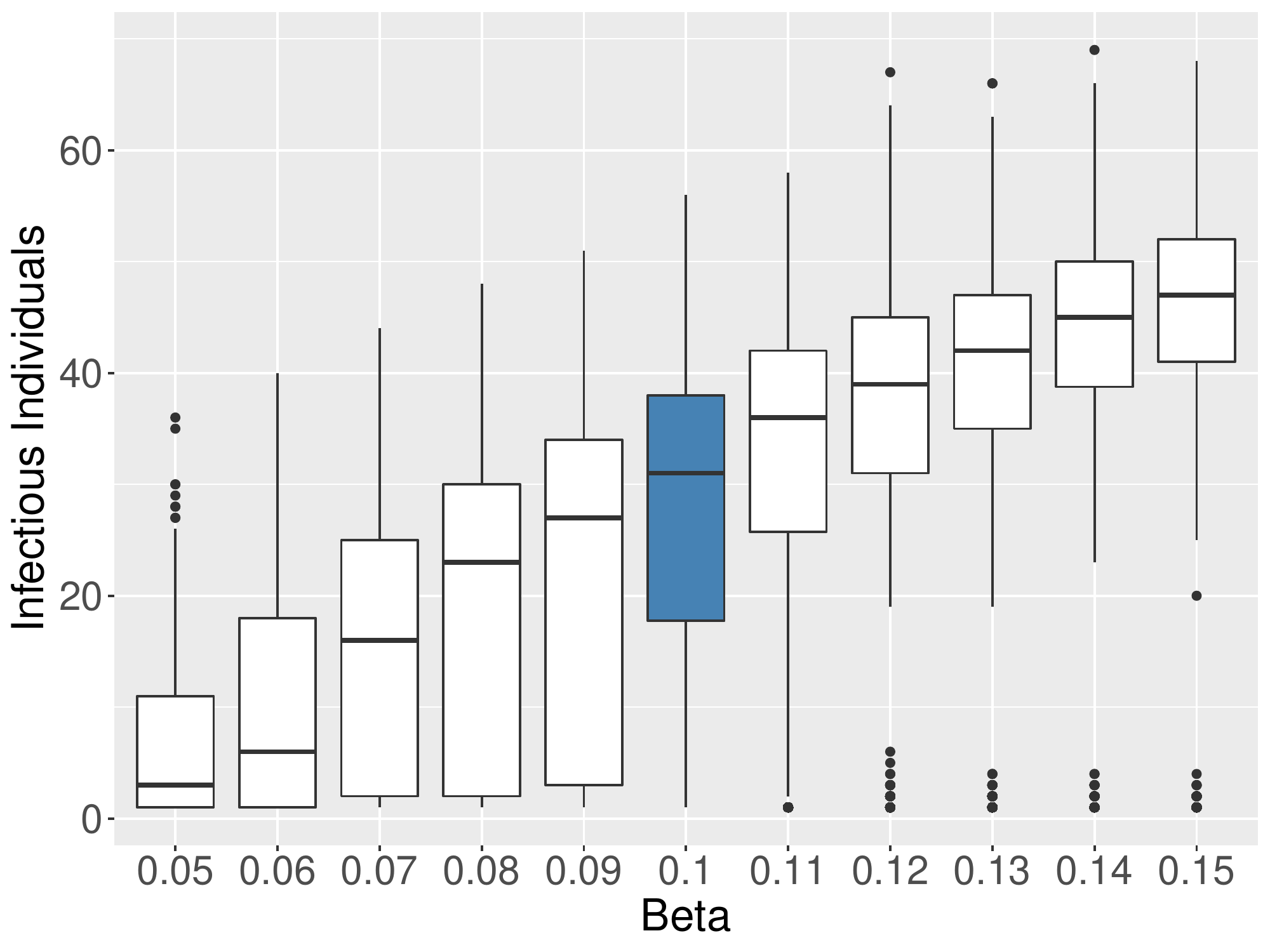}  
  \caption{Maximum Size of Infectious Class}
  \label{fig:Imaxbeta}
\end{subfigure}
\begin{subfigure}{0.5\textwidth}
  \centering
  \includegraphics[scale=0.45]{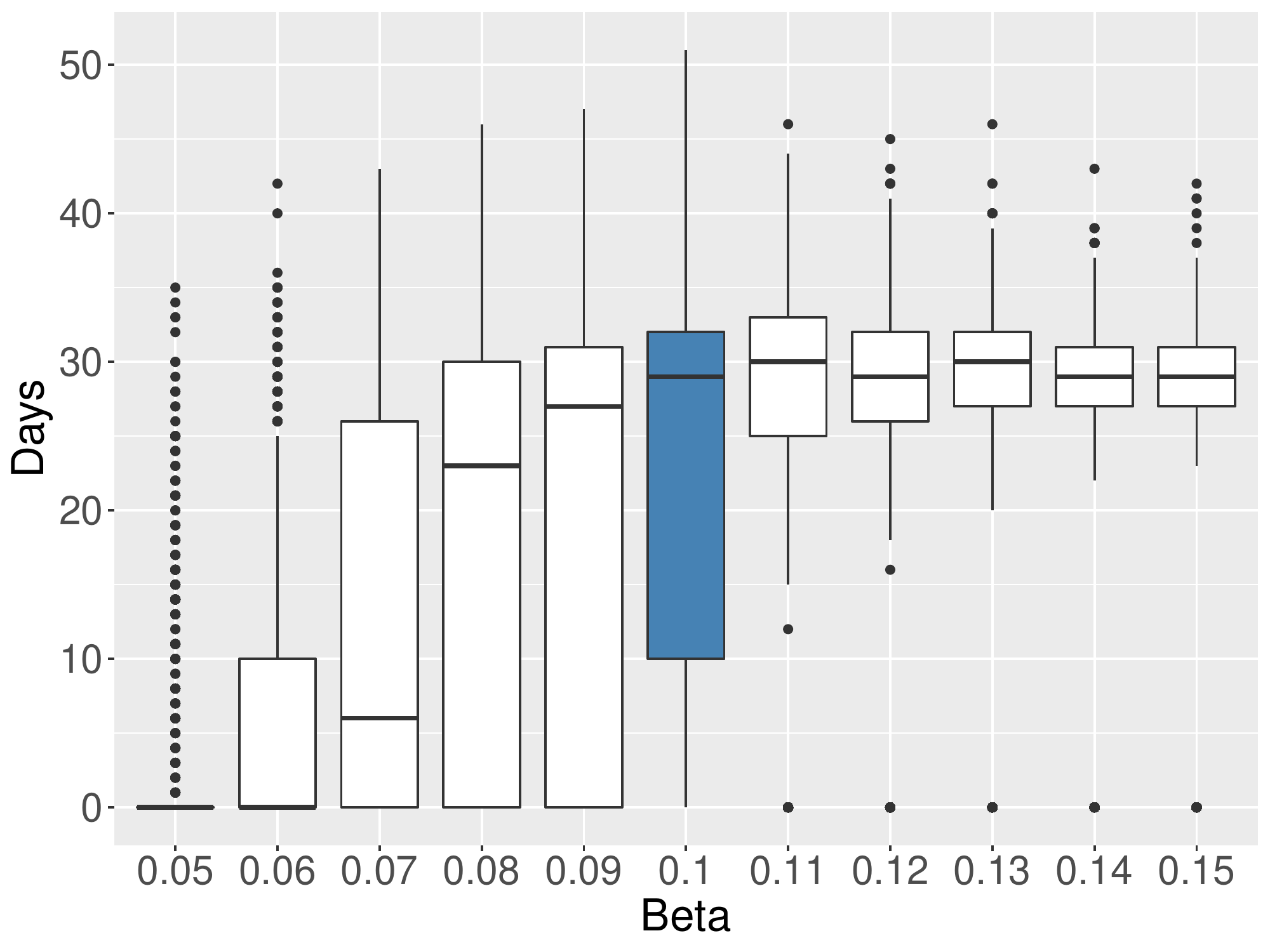}  
  \caption{Stress Days}
  \label{fig:stressbeta}
\end{subfigure}
\caption{Impact of epidemic on health care system for various values of the transmission rate.  Boxplots of the maximum size of the infectious class (a) and the number of stress days induced by the epidemic (b).  Values for the baseline parameter case are in blue.}
\label{fig:betaImaxstress}
\end{figure}

We can see that for the lowest values of the transmission rate, the epidemics are short, due to the fact that fewer people are infected.  The epidemic lengths rise rapidly with the lower values $\beta$, but taper off thereafter (for the larger values of $\beta$), as the faster transmission rates infect the population more rapidly.  The maximum size of the infectious group rises monotonically (nearly linearly) with the transmission rate (see Figure \ref{fig:Imaxbeta}).  However, our other indicator of the impact of the epidemic on the health care system, stress days, levels off at a median of about 30 days for $\beta \geq 0.1$ (see Figure \ref{fig:stressbeta}).  While higher transmission rates result in more individuals becoming infected, they also cause the epidemic to end (slightly) more rapidly; these two factors roughly offset to keep the number of stress days level for the larger values of $\beta.$

We also consider the values of the reproduction number for this disease, at the start and end of each simulated epidemic.  Table \ref{tab:R0beta} gives the median values of $R_0$ and $R_\omega$ for the simulated epidemics under the various values of the transmission rate.  The shapes of these distributions are similar to those seen in the baseline case; $R_0$ yields a symmetric, roughly bell-shaped distribution, whereas the distribution of $R_\omega$ is bimodal for the same reason as in the baseline case.

\begin{table}[h!t]
\centering
\begin{tabular}{lrrrrrrrrrrr}
  \hline
Transmission Rate & 0.05 & 0.06 & 0.07 & 0.08 & 0.09 & \textbf{0.10} & 0.11 & 0.12 & 0.13 & 0.14 & 0.15 \\ 
  \hline
$R_0$ (median) & 2.06 & 2.33 & 2.57 & 2.78 & 2.96 & \textbf{3.12} & 3.27 & 3.40 & 3.51 & 3.61 & 3.70 \\ 
  $R_\omega$ (median) & 1.90 & 1.99 & 1.44 & 1.34 & 1.34 & \textbf{1.33} & 1.34 & 1.35 & 1.37 & 1.41 & 1.40 \\ 
   \hline
\end{tabular}
\caption{Reproduction numbers at the start and end of simulated epidemics for various transmission rates.  Values for the baseline parameter case are bolded.} 
\label{tab:R0beta}
\end{table}

We can see that $R_0$ increases monotonically with $\beta$.  On the other hand, $R_\omega$ drops from values near 2 for the lowest transmission rates, to values near 1.4 for all $\beta \geq 0.07$.  In our formulation, the reproduction number is a function of both the degree distribution as well as the transmission rate (see Equation (\ref{eq:Rcalc})).  As the transmission rate increases, a greater proportion of the population ultimately becomes infected, and hence enters quarantine, reducing their contacts.  This has the effect of lowering their expected number of contacts, which largely offsets the direct effect that the increase in $\beta$ has on the reproduction number.

\subsubsection{Varying the Probability of Remaining Asymptomatic}
\label{subsubsec:asympt}

We next vary the probability that a given individual who becomes infected with the disease remains asymptomatic from values of $\alpha = 0.175$ to $\alpha = 0.525$.  As $\alpha$ increases, the average length of the epidemics changes little, and retains the same basic bimodal distribution shape as previously seen.  However, the dispersion of the length of the epidemics decreases as $\alpha$ increases; see Table \ref{tab:RmaxAprob} for summary statistics of the distributions of epidemic lengths under the various values of $\alpha$.


\begin{table}[h!t]
\centering
\begin{tabular}{lrrrrrrrrrrr}
  \hline
 $\alpha$ & 0.175 &0.210 &0.245 &0.280 &0.315 &\textbf{0.350} &0.385 &0.420 &0.455 &0.490 &0.525 \\ 
  \hline
Length (mean) & 69.8 & 70.0 & 69.7 & 69.5 & 69.7 &\textbf{70.0} & 70.5 & 71.7 & 71.3 & 70.9 & 70.9 \\ 
Length (st.dev.) & 37.1 & 36.9 & 35.8 & 34.0 & 33.3 & \textbf{32.1} & 31.0 & 31.1 & 30.1 & 28.8 & 28.0 \\ 
   \hline
\end{tabular}
\caption{Mean and standard deviation of lengths (in days) of simulated epidemics for various probabilities of remaining asymptomatic.  Values for the baseline parameter case are bolded. } 
\label{tab:RmaxAprob}
\end{table}

Next we consider the impact of varying $\alpha$ on the two metrics we use to assess the strain on the health care system imposed by the epidemic.  Table \ref{tab:healthAprob} gives the median values of these metrics across the various values of the asymptomatic probability.  We can see that in both cases, there is a roughly linear increase with $\alpha$, though we do note that the number of stress days levels off toward the higher end of the table.  Also, while the differences in these metrics across the values of $\alpha$ are significant, they are not particularly large in magnitude, compared to the changes in the parameter $\alpha$.

\begin{table}[h!t]
\centering
\begin{tabular}{lrrrrrrrrrrr}
  \hline
 $\alpha$ & 0.175 &0.210 &0.245 &0.280 &0.315 &\textbf{0.350} &0.385 &0.420 &0.455 &0.490 &0.525 \\ 
  \hline
\textbf{I}$_{max}$ (median) & 24 & 25 & 27 & 28 & 30 & \textbf{31} & 33 & 34 & 35 & 37 & 38 \\ 
$D_s$ (median) & 24 & 25 & 26 & 27 & 28 & \textbf{29} & 29 & 30 & 30 & 30 & 30 \\ 
   \hline
\end{tabular}
\caption{Impact of $\alpha$ on health care system strain.  Median values of the maximum size of the Infectious class and the number of stress days in simulated epidemics for various probabilities of remaining asymptomatic.  Values for the baseline parameter case are bolded.} 
\label{tab:healthAprob}
\end{table}

We again consider the values of the reproduction number for this disease, at the start and end of each simulated epidemic.  Table \ref{tab:R0Aprob} gives the median values of $R_0$ and $R_\omega$ for the simulated epidemics under the various values of $\alpha$.  The shapes of these distributions are similar to those seen previously; $R_0$ yields a symmetric, roughly bell-shaped distribution, whereas the distribution of $R_\omega$ is again bimodal.

\begin{table}[h!t]
\centering
\begin{tabular}{lrrrrrrrrrrr}
  \hline
$\alpha$ & 0.175 &0.210 &0.245 &0.280 &0.315 &\textbf{0.350} &0.385 &0.420 &0.455 &0.490 &0.525 \\ 
   \hline
$R_0$ (median) & 3.12 & 3.12 & 3.12 & 3.12 & 3.12 & \textbf{3.12} & 3.12 & 3.12 & 3.12 & 3.12 & 3.12 \\ 
$R_\omega$ (median) & 1.05 & 1.07 & 1.13 & 1.19 & 1.24 & \textbf{1.33} & 1.41 & 1.49 & 1.57 & 1.69 & 1.75 \\ 
   \hline
\end{tabular}
\caption{Reproduction numbers at the start and end of simulated epidemics for various probabilities of remaining asymptomatic.  Values for the baseline parameter case are bolded.} 
\label{tab:R0Aprob}
\end{table}

We note that the distribution of $R_0$ does not vary as the value of $\alpha$ changes.  In Equation (\ref{eq:Rcalc}), we see that $R$ is a function of the degree distribution; this degree distribution will change as individuals enter quarantine, but this does not occur until the epidemic is under way.  Hence, we should expect that $R_\omega$ will vary with $\alpha$, whereas $R_0$ will not, and this is indeed the case.  The median value of $R_0$, as noted earlier, is slightly higher than most other researchers have calculated or estimated, but is broadly reasonable and is suitable for our purposes.  With respect to $R_\omega$, it increases monotonically with $\alpha$, as we might expect.  Specifically, as the proportion of infected individuals who remain asymptomatic increases, fewer people enter quarantine.  Each infectious person who fails to enter quarantine continues to have unfettered opportunities to infect other individuals --- that is, the number of edges associated with the individual in the contact network does not decrease --- which prevents the reproduction number from dropping.

\subsubsection{Varying the Length of Time Spent in Exposed State}
\label{subsubsec:Etime}

To assess the effect of the length of time spent by individuals in the \textbf{E} state on the dynamics of this disease, we vary the mean of the (lognormal) distribution we use to model this time period from 1.45 days to 4.35 days; we adjust the standard deviation in each case in order to maintain a constant coefficient of variation.  As the mean length of time spent in the \textbf{E} state increases, the duration of the epidemics increases accordingly; this is quite intuitive, as individuals remaining latent for longer would be expected to lengthen the total duration of the epidemic.  The changes in epidemic duration are noticeable, but not dramatic.  This is also intuitive, considering that the baseline mean exposed period is relatively short (2.9 days).  Thus, increasing and decreasing this mean by 50\% should not be expected to make a large impact on the dynamics of the disease.  Table \ref{tab:RmaxE} gives the mean and median epidemic durations for the various mean exposure times.

\begin{table}[ht]
\centering
\begin{tabular}{lrrrrrrrrrrr}
  \hline
 Mean \textbf{E} Time & 1.45 &1.74& 2.03 &2.32 &2.61 &\textbf{2.90} &3.19& 3.48 &3.77 &4.06 &4.35 \\ 
  \hline
Length (mean) & 59 & 61 & 64 & 66 & 69 & \textbf{70} & 74 & 76 & 79 & 80 & 84 \\ 
Length (median) & 66 & 67 & 70 & 73 & 76 & \textbf{78} & 82 & 84 & 87 & 89 & 93 \\ 
   \hline
\end{tabular}
\caption{Mean and median lengths (in days) of simulated epidemics for various mean times in the Exposed state.  Values for the baseline parameter case are bolded.} 
\label{tab:RmaxE}
\end{table}

\begin{figure}[h!t]
\begin{subfigure}{0.5\textwidth}
  \centering
  \includegraphics[scale=0.45]{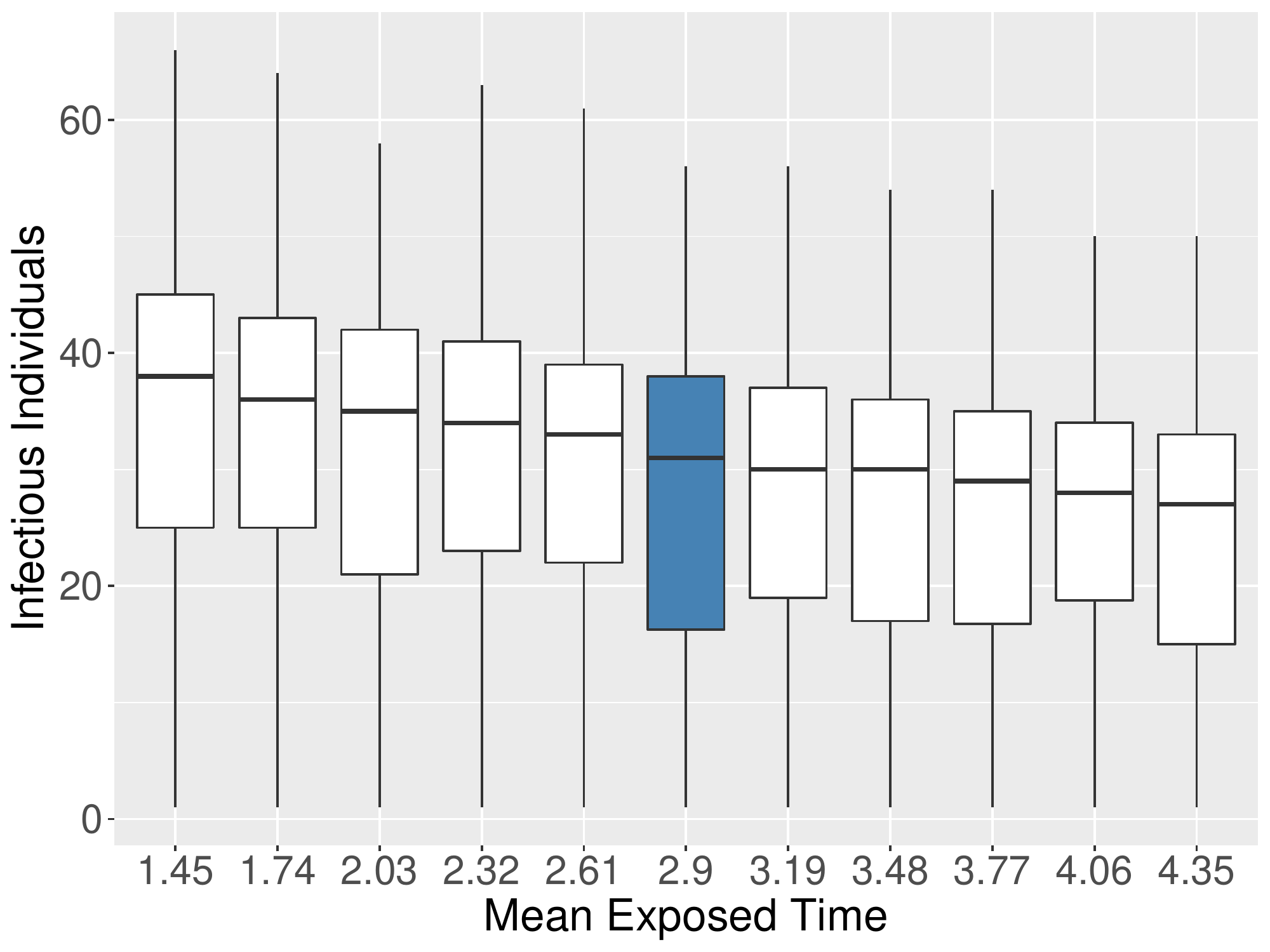}  
  \caption{Maximum Size of Infectious Class}
  \label{fig:ImaxE}
\end{subfigure}
\begin{subfigure}{0.5\textwidth}
  \centering
  \includegraphics[scale=0.45]{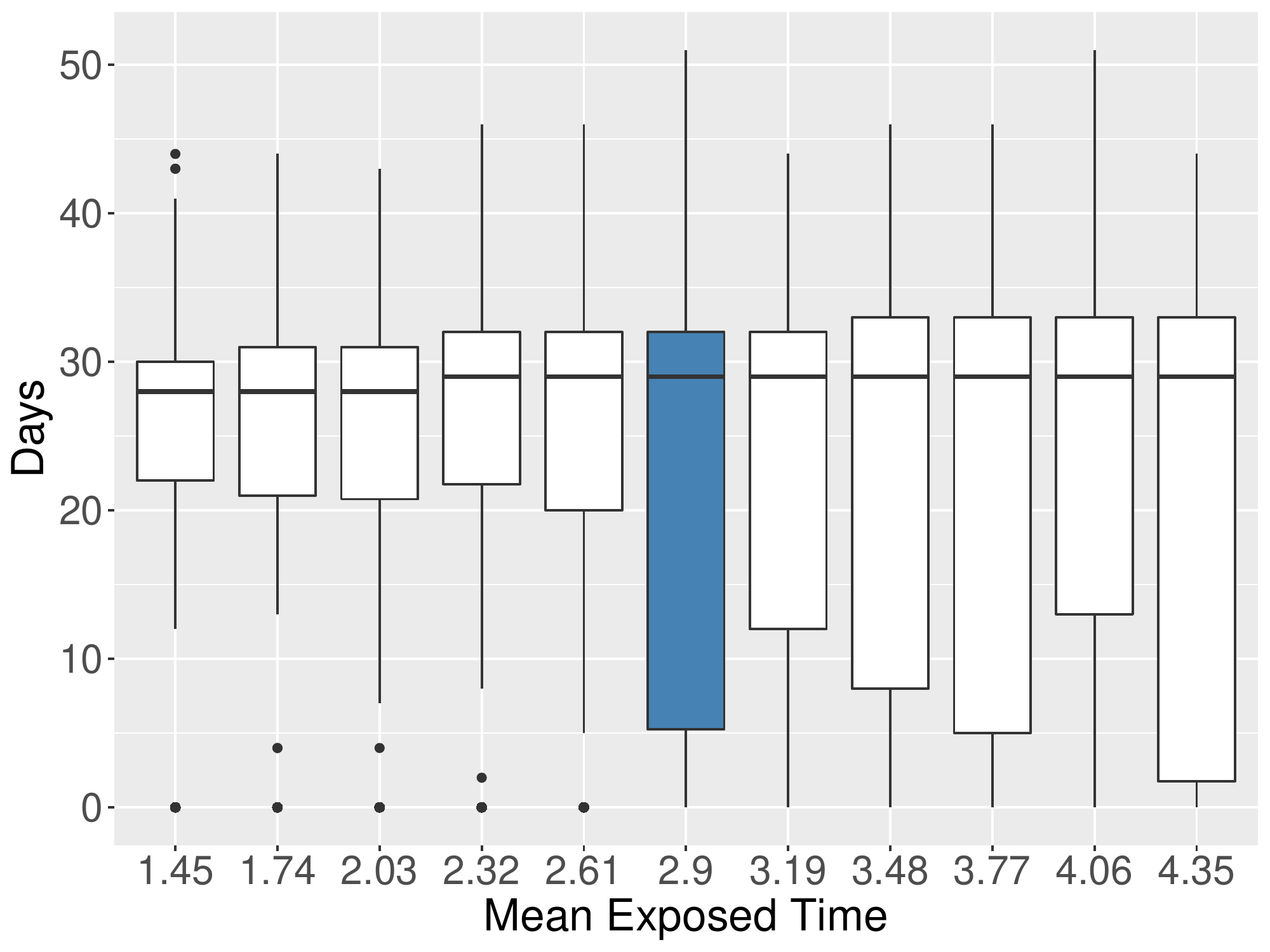}  
  \caption{Stress Days}
  \label{fig:stressEsub}
\end{subfigure}
\caption{Impact of epidemic on health care system for various values of the mean time spent in the \textbf{E} state.  Boxplots of the maximum size of the Infectious class (a) and the number of stress days induced by the epidemic (b).  Values for the baseline parameter case are in blue.}
\label{fig:stressE}
\end{figure}

We find that varying the length of time spent in the \textbf{E} does not tend to have a particularly great impact on the strain put on the health care system by the epidemic.  In particular, the maximum size of the Infectious class shrinks monotonically with the mean length of the exposed period; this is again due to the effect of spreading out the epidemic over a longer period of time (sometimes known as ``flattening the curve'').  The impact of mean \textbf{E} time on the number of stress days per epidemic is more subtle; while the median number of stress days stays roughly constant as the mean exposed time increases, there are an increasing number of epidemics with very few or no stress days, again a sign that the epidemic is being flattened.  Figure \ref{fig:stressE} shows boxplots for these two metrics as a function of the mean \textbf{E} time.

Based on our model and our methodology for calculating $R$, we would expect little or no impact on either $R_0$ or $R_\omega$ as a result of varying the mean time spent in the Exposed state.  Our results indicate that this is indeed the case, with the simulated distributions of both reproduction numbers staying very similar to those produced in the baseline case, regardless of the mean time spent in the \textbf{E} state; we see slightly more variability in $R_\omega$ than $R_0$, which conforms with our intuition, as the former metric reflects some additional variability in the empirical degree distribution due to the effects of quarantine.

\subsubsection{Varying the Length of Time Spent in Infectious State}
\label{subsubsec:Itime}

To assess the effect of the length of time spent by individuals in the \textbf{I} state on the dynamics of this disease, we vary the mean of the (gamma) distribution we use to model this time period from 6.25 days to 18.75 days; we adjust the standard deviation in each case in order to maintain a constant coefficient of variation.  As the mean length of time spent in the \textbf{I} state increases, the duration of the epidemics increases monotonically and roughly linearly.  The impact of the time spent in the Infectious state on the total length of the epidemic is somewhat greater than that for the Exposed state.  In addition to the direct impact on epidemic length of the changes in the \textbf{I} state, there is a secondary, indirect effect caused by an increase in the number of infectious individuals.  This latter effect occurs because when an individual spends a longer amount of time in the \textbf{I} state, they have more chances to infect others, thereby contributing to a lengthened epidemic duration.  The shape of the distribution of epidemic lengths remains bimodal, as in the other simulations.  Table \ref{tab:RmaxI} gives the mean and median epidemic lengths for the various mean infectious times.

\begin{table}[h!t]
\centering
\begin{tabular}{lrrrrrrrrrrr}
  \hline
 Mean \textbf{I} Time & 6.25 & 7.50 & 8.75 &10.00 &11.25 &\textbf{12.50} &13.75 &15.00& 16.25 &17.50 &18.75 \\ 
  \hline
Length (mean) & 47 & 54 & 58 & 62 & 66 & \textbf{70} & 74 & 78 & 81 & 84 & 86 \\ 
Length (median)& 52 & 62 & 67 & 70 & 74 & \textbf{78} & 81 & 85 & 87 & 90 & 92 \\ 
   \hline
\end{tabular}
\caption{Mean and median lengths (in days) of simulated epidemics for various mean times in the Infectious state.  Values for the baseline parameter case are bolded.} 
\label{tab:RmaxI}
\end{table}

\begin{figure}[h!t]
\begin{subfigure}{0.5\textwidth}
  \centering
  \includegraphics[scale=0.45]{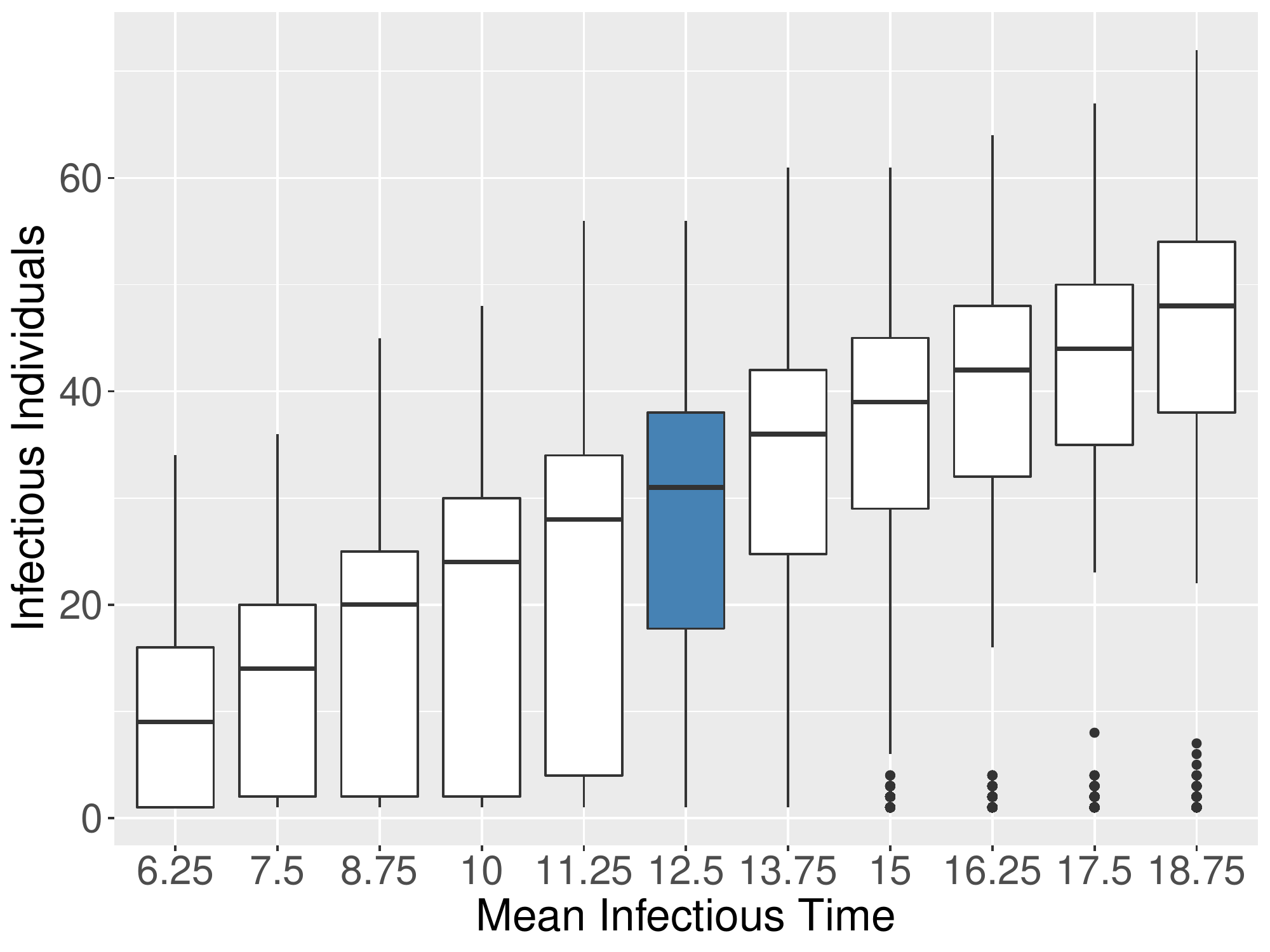}  
  \caption{Maximum Size of Infectious Class}
  \label{fig:ImaxI}
\end{subfigure}
\begin{subfigure}{0.5\textwidth}
  \centering
  \includegraphics[scale=0.45]{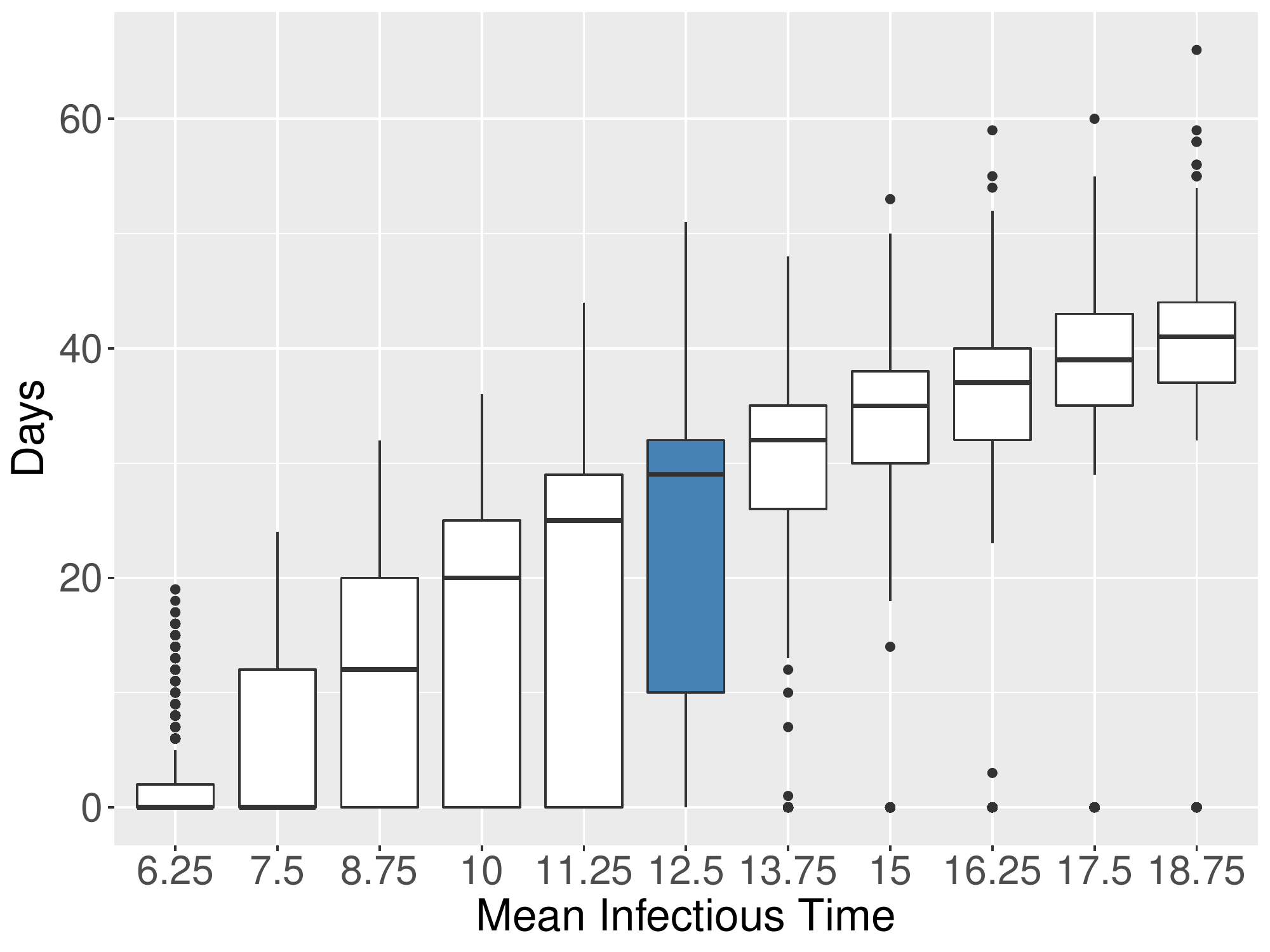}  
  \caption{Stress Days}
  \label{fig:stressdaysI}
\end{subfigure}
\caption{Impact of epidemic on health care system for various values of the mean time spent in the \textbf{I} state.  Boxplots of the maximum size of the Infectious class (a) and the number of stress days induced by the epidemic (b).  Values for the baseline parameter case are in blue.}
\label{fig:stressI}
\end{figure}

Unlike the previous section where we vary the length of time spent in the \textbf{E} state, we find that varying the length of time spent in the \textbf{I} state has a very significant impact on the strain put on the health care system by the epidemic.  The maximum size of the Infectious class increases substantially with the mean length of the \textbf{I} period; this is due both the direct effect of each individual remaining infectious for a longer period of time as well as an indirect effect as the result of a larger number of people becoming infected per epidemic.  These effects combine to produce resulting epidemics whose mean \textbf{I}$_{max}$ varies from 9.2 for the shortest mean infectious time up to 41.0 for the longest scenario.  Examining the number of stress days reveals a similar dynamic; this metric is impacted greatly by the changes in mean infectious time, with the mean number of stress days ranging from 2.1 days for the shortest mean infectious time up to 35.4 days for the longest mean infectious time.  Figure \ref{fig:stressI} shows boxplots for these two metrics as a function of the mean \textbf{I} time.

When we examine the reproduction number at the start of the epidemic, we see that it increases with the mean \textbf{I} length; this is expected, as under our model, the reproduction number is a function of the two parameters governing the length of time that each individual is infectious; see Equation (\ref{eq:Rcalc}).  Counterbalancing this effect in the calculation of $R_\omega$ (but not $R_0$) is the effect that longer periods of infectiousness lead to more individuals becoming infected.  This in turn leads to an increase in quarantined individuals, which decreases the mean degree in the underlying contact network.  The net result is that $R_\omega$ is relatively constant as a function of mean infectious period.  Table \ref{tab:R0I} gives the median vales of $R_0$ and $R_\omega$ for the various mean infectious lengths.

\begin{table}[h!t]
\centering
\begin{tabular}{lrrrrrrrrrrr}
  \hline
 Mean \textbf{I} Time & 6.25 & 7.50 & 8.75 &10.00 &11.25 &\textbf{12.50} &13.75 &15.00& 16.25 &17.50 &18.75 \\ 
  \hline
$R_0$ (median) & 2.06 & 2.33 & 2.57 & 2.78 & 2.96 & \textbf{3.12} & 3.27 & 3.40 & 3.51 & 3.61 & 3.70 \\ 
$R_{\omega}$ (median) & 1.45 & 1.29 & 1.31 & 1.30 & 1.31 & \textbf{1.33} & 1.35 & 1.37 & 1.38 & 1.41 & 1.46 \\ 
   \hline
\end{tabular}
\caption{Reproduction numbers at the start and end of simulated epidemics for various mean times in the Infectious state.  Values for the baseline parameter case are bolded.} 
\label{tab:R0I}
\end{table}

\subsubsection{Discussion}
\label{subsubsec:disc}

Because the SARS-CoV-2 virus is relatively new, it has yet to be studied in as much depth as many of the other viruses that commonly spread through populations.  However, as more researchers study the spread of this novel coronavirus, the understanding of its properties will continue to be refined, likely resulting in more definitive knowledge regarding model parameter values.  Our simulation study gives some indications about where this knowledge will be the most useful in shaping our understanding of the dynamics of this disease.

The transmission rate $\beta$ can be seen to have a significant impact on the dynamics of this disease, as seen in all of the metrics we calculate.  Namely, quicker transmission rates (greater values of $\beta$) tend to lead to epidemics having longer durations and ultimately infecting greater numbers of people.  We do note, however, that some of the metrics, in particular, the number of stress days and $R_\omega$ level off for the larger values of $\beta$.  That is, the epidemics are sensitive to the value of $\beta$ for values of this parameter less than about 0.1, but relatively insensitive (by most metrics) to changes in the transmission rate above this point.  We also note that in our model infectiousness is a binary property; some other researchers allow for different levels of infectiousness as a function of time, e.g., \citet{lin2020conceptual}.  We do not feel the current body of research provides sufficient information to accurately model this; improving knowledge of this virus may provide guidance in this area.

Increasing the probability that an individual remains asymptomatic throughout the course of the disease led to more severe epidemics by all of the metrics we examined.  This is a result of fewer individuals quarantining, and hence reducing their contacts, which would lead to fewer infections.  However, while the effect of the parameter $\alpha$ is monotonic, the magnitude of the effects --- as measured by all of our metrics ---  is comparatively small.  Thus, it appears that more precise knowledge of this parameter may not lead to great changes in the epidemic outcomes for this disease.

As the mean length of time spent in the Exposed state increases, the trajectory of the disease tends to be elongated somewhat.  That is, the total length of the disease in increased, while the maximum size of the Infectious class decreases.  We note that these effects are relatively small in magnitude, and that the median number of stress days is mostly unchanged as the mean \textbf{E} time increases.  We can conclude that the epidemics involving this disease are relatively insensitive to this particular parameter.

Finally, we see that the length of time spent in the Infectious class makes a very significant impact on the dynamics of this disease.  With the exception of $R_\omega$, all of the metrics we calculate indicate that the severity of an epidemic is quite sensitive to the length of time spent in the \textbf{I} state.  Hence, it seems likely that better knowledge about this model parameter is very likely to lead to significantly more accurate modeling of the spread of this disease through populations.  

\section{Conclusions}
\label{sec:conc}

The SARS-CoV-2 virus and accompanying COVID-19 pandemic has had an enormous impact across the globe, disrupting many aspects of society, from health outcomes to individual behavior to financial markets.  With no effective treatments having yet been discovered, vaccines not yet being available, and the possibility of new strains of this virus circulating, it is clearly important to work to improve our understanding of the dynamics of this disease.  Our work furthers this effort by presenting a viable alternative to the mean field model, which has been by far the most utilized framework in the early studies of this pandemic.

In this paper, we have presented a modification to the network-based stochastic SEIR model to account for the effects of quarantine or self-isolation on the contact patterns of individuals in a population.  Our model also allows us to incorporate the effects of a percentage of the population remaining asymptomatic during the entirety of their time in the Infectious state.  Both of these model features are important to the accurate modeling of the SARS-CoV-2 virus.  

Using a GER model for the underlying contact network, our simulation study examined the sensitivity of the severity of epidemics to four basic model parameters: transmission rate, probability of remaining asymptomatic, and the mean lengths of time spent in the \textbf{E} and \textbf{I} states.  We found that the length of time spent in the Infectious state was the most important driver of the epidemic severity and that the transmission rate was also important; most metrics were far less sensitive to changes in the time spend in the Exposed state and the probability of remaining asymptomatic.  The metrics examined in the study included the total duration of the epidemic, the maximum size of the Infectious class and number of stress days (as proxies for strain on the health care system), and the estimated reproduction number at the start and end of the epidemic. 

Our application of this model to the novel coronavirus assumes that individuals will self-isolate or quarantine at the onset of disease symptoms, which will tend to reduce their number of connections to other individuals.  We note that our framework is more general, though, and could be used to model any situation where an individual might change their contact patterns, either by adding or removing contacts.

There are several potential directions for future research based on this model, especially in terms of extensions of the model and applications to various types of populations.  While we have used the simple GER model to describe the underlying contact network for this study, it has been previously demonstrated that accounting for more accurate contact network patterns can produce more realistic disease dynamics.  Hence, an obvious avenue for future work would be to consider the spread of this virus through specific types of populations, incorporating a non-trivial network model.  One possibility would be to utilize a type of Exponential-family Random Graph Model (ERGM) or $p^\ast$ model \citep{holland1981exponential, wasserman1996logit}.  This family of models is very flexible (allowing for dyadic dependence or independence), well-studied \citep{robins2007introduction, hunter2008ergm}, and contains some important special cases, such as the GER model.

Due to the explicit modeling of each individual's contacts, the framework presented here is most useful for relatively small and closed populations.  While the modeling of a country or state is likely not feasible, there are important sub-populations that would likely fit well into this framework; some possibilities include settings like nursing homes, small colleges, prisons, or cruise ships.  Our model not only allows researchers to simulate the spread of epidemics through such populations, but also enables explicit testing of various containment strategies that might be implemented in such sub-populations.  This type of study would seem to be both important and timely.

\section*{Acknowledgments}

The authors would like to acknowledge David Welch, who wrote the original code in the {\tt epinet} package \citep{groendyke2018epinet} for simulating an epidemic using a stochastic network-based SEIR model; the code we used in this study was based on this work.

\bibliographystyle{chicagoa}
\bibliography{seiqrbib}

\begin{thebibliography}{}

\bibitem[\protect\citeauthoryear{Anderson and May}{Anderson and
  May}{1992}]{anderson1992infectious}
Anderson, R.~M. and R.~M. May (1992).
\newblock {\em Infectious diseases of humans: dynamics and control}.
\newblock Oxford University Press.


\bibitem[\protect\citeauthoryear{Andersson}{Andersson}{1998}]{andersson1998limit}
Andersson, H. (1998).
\newblock Limit theorems for a random graph epidemic model.
\newblock {\em Annals of Applied Probability\/}, 1331--1349.


\bibitem[\protect\citeauthoryear{Aron and Schwartz}{Aron and
  Schwartz}{1984}]{aron1984seasonality}
Aron, J.~L. and I.~B. Schwartz (1984).
\newblock Seasonality and period-doubling bifurcations in an epidemic model.
\newblock {\em Journal of Theoretical Biology\/}~{\em 110\/}(4), 665--679.


\bibitem[\protect\citeauthoryear{Bailey}{Bailey}{1950}]{bailey1950sse}
Bailey, N. (1950).
\newblock {A simple stochastic epidemic}.
\newblock {\em Biometrika\/}~{\em 37\/}(3-4), 193--202.


\bibitem[\protect\citeauthoryear{Bansal, Read, Pourbohloul, and Meyers}{Bansal
  et~al.}{2010}]{bansal2010dynamic}
Bansal, S., J.~Read, B.~Pourbohloul, and L.~A. Meyers (2010).
\newblock The dynamic nature of contact networks in infectious disease
  epidemiology.
\newblock {\em Journal of Biological Dynamics\/}~{\em 4\/}(5), 478--489.


\bibitem[\protect\citeauthoryear{Barthelemy, Barrat, Pastor-Satorras, and
  Vespignani}{Barthelemy et~al.}{2005}]{barthelemy2005dpe}
Barthelemy, M., A.~Barrat, R.~Pastor-Satorras, and A.~Vespignani (2005).
\newblock {Dynamical patterns of epidemic outbreaks in complex heterogeneous
  networks}.
\newblock {\em Journal of Theoretical Biology\/}~{\em 235\/}(2), 275--288.


\bibitem[\protect\citeauthoryear{{Center for Disease Control and
  Prevention}}{{Center for Disease Control and Prevention}}{2020}]{cdc2020}
{Center for Disease Control and Prevention} (2020).
\newblock {COVID-19} pandemic planning scenarios.
\newblock
  \url{https://www.cdc.gov/coronavirus/2019-ncov/hcp/planning-scenarios.html#table-2}.
\newblock Accessed: 2020-06-22.

\bibitem[\protect\citeauthoryear{Dukic, Lopes, and Polson}{Dukic
  et~al.}{2012}]{dukic2012tracking}
Dukic, V., H.~F. Lopes, and N.~G. Polson (2012).
\newblock Tracking epidemics with {G}oogle flu trends data and a state-space
  {SEIR} model.
\newblock {\em Journal of the American Statistical Association\/}~{\em
  107\/}(500), 1410--1426.


\bibitem[\protect\citeauthoryear{Erd\H{o}s and R\'enyi}{Erd\H{o}s and
  R\'enyi}{1959}]{erdos1959rg}
Erd\H{o}s, P. and A.~R\'enyi (1959).
\newblock {On random graphs}.
\newblock {\em Publ. Math. Debrecen\/}~{\em 6\/}(290).


\bibitem[\protect\citeauthoryear{Fang, Nie, and Penny}{Fang
  et~al.}{2020}]{fang2020transmission}
Fang, Y., Y.~Nie, and M.~Penny (2020).
\newblock Transmission dynamics of the {COVID-19} outbreak and effectiveness of
  government interventions: {A} data-driven analysis.
\newblock {\em Journal of Medical Virology\/}~{\em 92\/}(6), 645--659.


\bibitem[\protect\citeauthoryear{Ferrari, Bansal, Meyers, and
  Bj{\o}rnstad}{Ferrari et~al.}{2006}]{ferrari2006network}
Ferrari, M.~J., S.~Bansal, L.~A. Meyers, and O.~N. Bj{\o}rnstad (2006).
\newblock Network frailty and the geometry of herd immunity.
\newblock {\em Proceedings of the Royal Society B: Biological Sciences\/}~{\em
  273\/}(1602), 2743--2748.


\bibitem[\protect\citeauthoryear{Gilbert}{Gilbert}{1959}]{gilbert1959random}
Gilbert, E. (1959).
\newblock {Random graphs}.
\newblock {\em The Annals of Mathematical Statistics\/}, 1141--1144.


\bibitem[\protect\citeauthoryear{Gonz{\'a}lez-Parra, Arenas, and
  Chen-Charpentier}{Gonz{\'a}lez-Parra et~al.}{2014}]{gonzalez2014fractional}
Gonz{\'a}lez-Parra, G., A.~J. Arenas, and B.~M. Chen-Charpentier (2014).
\newblock A fractional order epidemic model for the simulation of outbreaks of
  influenza {A} ({H1N1}).
\newblock {\em Mathematical Methods in the Applied Sciences\/}~{\em 37\/}(15),
  2218--2226.


\bibitem[\protect\citeauthoryear{Grais, Ellis, and Glass}{Grais
  et~al.}{2003}]{grais2003assessing}
Grais, R.~F., J.~H. Ellis, and G.~E. Glass (2003).
\newblock Assessing the impact of airline travel on the geographic spread of
  pandemic influenza.
\newblock {\em European Journal of Epidemiology\/}~{\em 18\/}(11), 1065--1072.


\bibitem[\protect\citeauthoryear{Grenfell}{Grenfell}{1992}]{grenfell1992chance}
Grenfell, B. (1992).
\newblock Chance and chaos in measles dynamics.
\newblock {\em Journal of the Royal Statistical Society: Series B
  (Methodological)\/}~{\em 54\/}(2), 383--398.


\bibitem[\protect\citeauthoryear{Groendyke and Welch}{Groendyke and
  Welch}{2018}]{groendyke2018epinet}
Groendyke, C. and D.~Welch (2018).
\newblock epinet: An {R} package to analyze epidemics spread across contact
  networks.
\newblock {\em Journal of Statistical Software\/}~{\em 83\/}(11), 1--22.


\bibitem[\protect\citeauthoryear{Groendyke, Welch, and Hunter}{Groendyke
  et~al.}{2012}]{groendyke2012network}
Groendyke, C., D.~Welch, and D.~R. Hunter (2012).
\newblock A network-based analysis of the 1861 {H}agelloch measles data.
\newblock {\em Biometrics\/}~{\em 68\/}(3), 755--765.


\bibitem[\protect\citeauthoryear{He, Lau, Wu, and et~al.}{He
  et~al.}{2020}]{He2020NatureMed}
He, X., E.~Lau, P.~Wu, and et~al. (2020).
\newblock Temporal dynamics in viral shedding and transmissibility of{
  COVID-19}.
\newblock {\em Nature Medicine\/}~{\em 26}, 672--675.
\newblock doi:10.1038/s41591-020-0869-5.


\bibitem[\protect\citeauthoryear{Hethcote and Tudor}{Hethcote and
  Tudor}{1980}]{hethcote1980integral}
Hethcote, H.~W. and D.~W. Tudor (1980).
\newblock Integral equation models for endemic infectious diseases.
\newblock {\em Journal of Mathematical Biology\/}~{\em 9\/}(1), 37--47.


\bibitem[\protect\citeauthoryear{Holland and Leinhardt}{Holland and
  Leinhardt}{1981}]{holland1981exponential}
Holland, P. and S.~Leinhardt (1981).
\newblock {An exponential family of probability distributions for directed
  graphs}.
\newblock {\em Journal of the American Statistical Association\/}, 33--50.


\bibitem[\protect\citeauthoryear{Hou, Chen, Zhou, Hua, Yuan, He, Guo, Zhang,
  Jia, Zhao, et~al.}{Hou et~al.}{2020}]{hou2020effectiveness}
Hou, C., J.~Chen, Y.~Zhou, L.~Hua, J.~Yuan, S.~He, Y.~Guo, S.~Zhang, Q.~Jia,
  C.~Zhao, et~al. (2020).
\newblock The effectiveness of quarantine of {W}uhan city against the corona
  virus disease 2019 ({COVID-19}): A well-mixed {SEIR} model analysis.
\newblock {\em Journal of Medical Virology\/}.


\bibitem[\protect\citeauthoryear{Hunter, Handcock, Butts, Goodreau, and
  Morris}{Hunter et~al.}{2008}]{hunter2008ergm}
Hunter, D.~R., M.~S. Handcock, C.~T. Butts, S.~M. Goodreau, and M.~Morris
  (2008).
\newblock ergm: A package to fit, simulate and diagnose exponential-family
  models for networks.
\newblock {\em Journal of Statistical Software\/}~{\em 24\/}(3), nihpa54860.


\bibitem[\protect\citeauthoryear{Iwata and Miyakoshi}{Iwata and
  Miyakoshi}{2020}]{iwata2020simulation}
Iwata, K. and C.~Miyakoshi (2020).
\newblock A simulation on potential secondary spread of novel coronavirus in an
  exported country using a stochastic epidemic {SEIR} model.
\newblock {\em Journal of Clinical Medicine\/}~{\em 9\/}(4), 944.


\bibitem[\protect\citeauthoryear{Keeling and Eames}{Keeling and
  Eames}{2005}]{keeling2005nae}
Keeling, M. and K.~Eames (2005).
\newblock {Networks and epidemic models}.
\newblock {\em Journal of the Royal Society Interface\/}~{\em 2\/}(4), 295.


\bibitem[\protect\citeauthoryear{Keeling and Rohani}{Keeling and
  Rohani}{2011}]{keeling2011modeling}
Keeling, M.~J. and P.~Rohani (2011).
\newblock {\em Modeling infectious diseases in humans and animals}.
\newblock Princeton University Press.


\bibitem[\protect\citeauthoryear{Kenah}{Kenah}{2011}]{kenah2011contact}
Kenah, E. (2011).
\newblock Contact intervals, survival analysis of epidemic data, and estimation
  of $r_0$.
\newblock {\em Biostatistics\/}~{\em 12\/}(3), 548--566.


\bibitem[\protect\citeauthoryear{Kermack and McKendrick}{Kermack and
  McKendrick}{1927}]{kermack1927cmt}
Kermack, W. and A.~McKendrick (1927).
\newblock {A Contribution to the Mathematical Theory of Epidemics}.
\newblock {\em Proceedings of the Royal Society of London. Series A, Containing
  Papers of a Mathematical and Physical Character (1905-1934)\/}~{\em
  115\/}(772), 700--721.


\bibitem[\protect\citeauthoryear{Kuniya}{Kuniya}{2020}]{kuniya2020prediction}
Kuniya, T. (2020).
\newblock Prediction of the epidemic peak of coronavirus disease in {J}apan,
  2020.
\newblock {\em Journal of Clinical Medicine\/}~{\em 9\/}(3), 789.


\bibitem[\protect\citeauthoryear{Li, Smith, and Wang}{Li
  et~al.}{2001}]{li2001global}
Li, M.~Y., H.~L. Smith, and L.~Wang (2001).
\newblock Global dynamics of an {SEIR} epidemic model with vertical
  transmission.
\newblock {\em SIAM Journal on Applied Mathematics\/}~{\em 62\/}(1), 58--69.


\bibitem[\protect\citeauthoryear{Li, Guan, Wu, and Wang}{Li
  et~al.}{2020}]{Li2020NEJM}
Li, Q., X.~Guan, P.~Wu, and X.~Wang (2020).
\newblock Early transmission dynamics in {W}uhan, {C}hina, of novel
  coronavirus-infected pneumonia.
\newblock {\em The New England Journal of Medicine\/}~{\em 382\/}(13),
  1199--1207.


\bibitem[\protect\citeauthoryear{Lin, Zhao, Gao, Lou, Yang, Musa, Wang, Cai,
  Wang, Yang, et~al.}{Lin et~al.}{2020}]{lin2020conceptual}
Lin, Q., S.~Zhao, D.~Gao, Y.~Lou, S.~Yang, S.~S. Musa, M.~H. Wang, Y.~Cai,
  W.~Wang, L.~Yang, et~al. (2020).
\newblock A conceptual model for the outbreak of {C}oronavirus disease 2019
  ({COVID-19}) in {W}uhan, {C}hina with individual reaction and governmental
  action.
\newblock {\em International Journal of Infectious Diseases\/}.


\bibitem[\protect\citeauthoryear{Ling, Xu, Lin, Tian, Zhu, Dai, Wu, Song,
  Huang, Chen, et~al.}{Ling et~al.}{2020}]{ling2020persistence}
Ling, Y., S.-B. Xu, Y.-X. Lin, D.~Tian, Z.-Q. Zhu, F.-H. Dai, F.~Wu, Z.-G.
  Song, W.~Huang, J.~Chen, et~al. (2020).
\newblock Persistence and clearance of viral {RNA} in 2019 novel coronavirus
  disease rehabilitation patients.
\newblock {\em Chinese Medical Journal\/}.


\bibitem[\protect\citeauthoryear{Linton, Kobayashi, Yang, and Hayashi}{Linton
  et~al.}{2020}]{Linton2020JCM}
Linton, N., T.~Kobayashi, Y.~Yang, and K.~Hayashi (2020).
\newblock Incubation period and other epidemiological characteristics of 2019
  novel coronavirus infections with right truncation: a statistical analysis of
  publicly available case data.
\newblock {\em Journal of Clinical Medicine\/}~{\em 9\/}(2).
\newblock doi:10.3390/jcm9020538.


\bibitem[\protect\citeauthoryear{Liu, Gayle, Wilder-Smith, and Rockl{\"o}v}{Liu
  et~al.}{2020}]{liu2020reproductive}
Liu, Y., A.~A. Gayle, A.~Wilder-Smith, and J.~Rockl{\"o}v (2020).
\newblock The reproductive number of {COVID-19} is higher compared to {SARS}
  coronavirus.
\newblock {\em Journal of Travel Medicine\/}.


\bibitem[\protect\citeauthoryear{L{\'o}pez and Rodo}{L{\'o}pez and
  Rodo}{2020}]{lopez2020modified}
L{\'o}pez, L. and X.~Rodo (2020).
\newblock A modified {SEIR} model to predict the {COVID-19} outbreak in {S}pain
  and {I}taly: simulating control scenarios and multi-scale epidemics.
\newblock {\em Available at SSRN 3576802\/}.


\bibitem[\protect\citeauthoryear{Meyers}{Meyers}{2007}]{meyers2007contact}
Meyers, L. (2007).
\newblock Contact network epidemiology: Bond percolation applied to infectious
  disease prediction and control.
\newblock {\em Bulletin of the American Mathematical Society\/}~{\em 44\/}(1),
  63--86.


\bibitem[\protect\citeauthoryear{Meyers, Pourbohloul, Newman, Skowronski, and
  Brunham}{Meyers et~al.}{2005}]{meyers2005nta}
Meyers, L., B.~Pourbohloul, M.~Newman, D.~Skowronski, and R.~Brunham (2005).
\newblock {Network theory and SARS: predicting outbreak diversity}.
\newblock {\em Journal of Theoretical Biology\/}~{\em 232\/}(1), 71--81.


\bibitem[\protect\citeauthoryear{Momoh, Ibrahim, Uwanta, and Manga}{Momoh
  et~al.}{2013}]{momoh2013mathematical}
Momoh, A., M.~Ibrahim, I.~Uwanta, and S.~Manga (2013).
\newblock Mathematical model for control of measles epidemiology.
\newblock {\em International Journal of Pure and Applied Mathematics\/}~{\em
  87\/}(5), 707--717.


\bibitem[\protect\citeauthoryear{Nishiura, Kobayashi, Miyama, Suzuki, Jung,
  Hayashi, Kinoshita, Yang, Yuan, Akhmetzhanov, et~al.}{Nishiura
  et~al.}{2020}]{nishiura2020estimation}
Nishiura, H., T.~Kobayashi, T.~Miyama, A.~Suzuki, S.-m. Jung, K.~Hayashi,
  R.~Kinoshita, Y.~Yang, B.~Yuan, A.~R. Akhmetzhanov, et~al. (2020).
\newblock Estimation of the asymptomatic ratio of novel coronavirus infections
  ({COVID-19}).
\newblock {\em International Journal of Infectious Diseases\/}~{\em 94}, 154.


\bibitem[\protect\citeauthoryear{Nkwayep, Bowong, Tewa, and Kurths}{Nkwayep
  et~al.}{2020}]{nkwayep2020short}
Nkwayep, C.~H., S.~Bowong, J.~Tewa, and J.~Kurths (2020).
\newblock Short-term forecasts of the {COVID-19} pandemic: study case of
  {C}ameroon.
\newblock {\em Chaos, Solitons \& Fractals\/}, 110106.


\bibitem[\protect\citeauthoryear{Peng, Yang, Zhang, Zhuge, and Hong}{Peng
  et~al.}{2020}]{peng2020epidemic}
Peng, L., W.~Yang, D.~Zhang, C.~Zhuge, and L.~Hong (2020).
\newblock Epidemic analysis of {COVID-19} in {C}hina by dynamical modeling.
\newblock {\em arXiv preprint arXiv:2002.06563\/}.


\bibitem[\protect\citeauthoryear{Prasse, Achterberg, Ma, and
  Van~Mieghem}{Prasse et~al.}{2020}]{prasse2020network}
Prasse, B., M.~A. Achterberg, L.~Ma, and P.~Van~Mieghem (2020).
\newblock Network-based prediction of the {2019-nCoV} epidemic outbreak in the
  {C}hinese province {H}ubei.
\newblock {\em arXiv preprint arXiv:2002.04482\/}.


\bibitem[\protect\citeauthoryear{Radulescu and Cavanagh}{Radulescu and
  Cavanagh}{2020}]{radulescu2020management}
Radulescu, A. and K.~Cavanagh (2020).
\newblock Management strategies in a {SEIR} model of {COVID-19} community
  spread.
\newblock {\em arXiv preprint arXiv:2003.11150\/}.


\bibitem[\protect\citeauthoryear{Robins, Pattison, Kalish, and Lusher}{Robins
  et~al.}{2007}]{robins2007introduction}
Robins, G., P.~Pattison, Y.~Kalish, and D.~Lusher (2007).
\newblock An introduction to exponential random graph (p*) models for social
  networks.
\newblock {\em Social networks\/}~{\em 29\/}(2), 173--191.


\bibitem[\protect\citeauthoryear{Schwartz and Smith}{Schwartz and
  Smith}{1983}]{schwartz1983infinite}
Schwartz, I.~B. and H.~Smith (1983).
\newblock Infinite subharmonic bifurcation in an {SEIR} epidemic model.
\newblock {\em Journal of Mathematical Biology\/}~{\em 18\/}(3), 233--253.


\bibitem[\protect\citeauthoryear{Shi, Cao, and Feng}{Shi
  et~al.}{2020}]{shi2020seir}
Shi, P., S.~Cao, and P.~Feng (2020).
\newblock {SEIR} transmission dynamics model of 2019 {nCoV} coronavirus with
  considering the weak infectious ability and changes in latency duration.
\newblock {\em medRxiv\/}.


\bibitem[\protect\citeauthoryear{{The New York Times}}{{The New York
  Times}}{2020}]{nyt2020}
{The New York Times} (2020).
\newblock Coronavirus map: Tracking the global outbreak.
\newblock
  \url{https://www.nytimes.com/interactive/2020/world/coronavirus-maps.html}.
\newblock Accessed: 2020-06-17.

\bibitem[\protect\citeauthoryear{Volz}{Volz}{2008}]{volz2008sdr}
Volz, E. (2008).
\newblock {SIR dynamics in random networks with heterogeneous connectivity}.
\newblock {\em Journal of Mathematical Biology\/}~{\em 56\/}(3), 293--310.


\bibitem[\protect\citeauthoryear{Wan, Chen, Lu, Dong, Wu, and Zhang}{Wan
  et~al.}{2020}]{wan2020will}
Wan, K., J.~Chen, C.~Lu, L.~Dong, Z.~Wu, and L.~Zhang (2020).
\newblock When will the battle against novel coronavirus end in {W}uhan: A
  {SEIR} modeling analysis.
\newblock {\em Journal of Global Health\/}~{\em 10\/}(1).


\bibitem[\protect\citeauthoryear{Wang, Liu, Hao, Guo, Wang, Huang, He, Yu, Lin,
  Pan, et~al.}{Wang et~al.}{2020}]{wang2020evolving}
Wang, C., L.~Liu, X.~Hao, H.~Guo, Q.~Wang, J.~Huang, N.~He, H.~Yu, X.~Lin,
  A.~Pan, et~al. (2020).
\newblock Evolving epidemiology and impact of non-pharmaceutical interventions
  on the outbreak of {C}oronavirus disease 2019 in {W}uhan, {C}hina.
\newblock {\em MedRxiv\/}.


\bibitem[\protect\citeauthoryear{Wasserman and Pattison}{Wasserman and
  Pattison}{1996}]{wasserman1996logit}
Wasserman, S. and P.~Pattison (1996).
\newblock Logit models and logistic regressions for social networks: I. an
  introduction to {M}arkov graphs and {$p^\ast$}.
\newblock {\em Psychometrika\/}~{\em 61\/}(3), 401--425.


\bibitem[\protect\citeauthoryear{Wearing, Rohani, and Keeling}{Wearing
  et~al.}{2005}]{wearing2005appropriate}
Wearing, H.~J., P.~Rohani, and M.~J. Keeling (2005).
\newblock Appropriate models for the management of infectious diseases.
\newblock {\em PLoS Medicine\/}~{\em 2\/}(7).


\end{thebibliography}

\newpage
\begin{appendices}
\section{Transmission Tree for Sample Simulated Epidemic}
\label{sec:app}

\begin{figure}[h!t]
  \centering
  \includegraphics[scale=0.91, trim=42 2 12 42, clip]{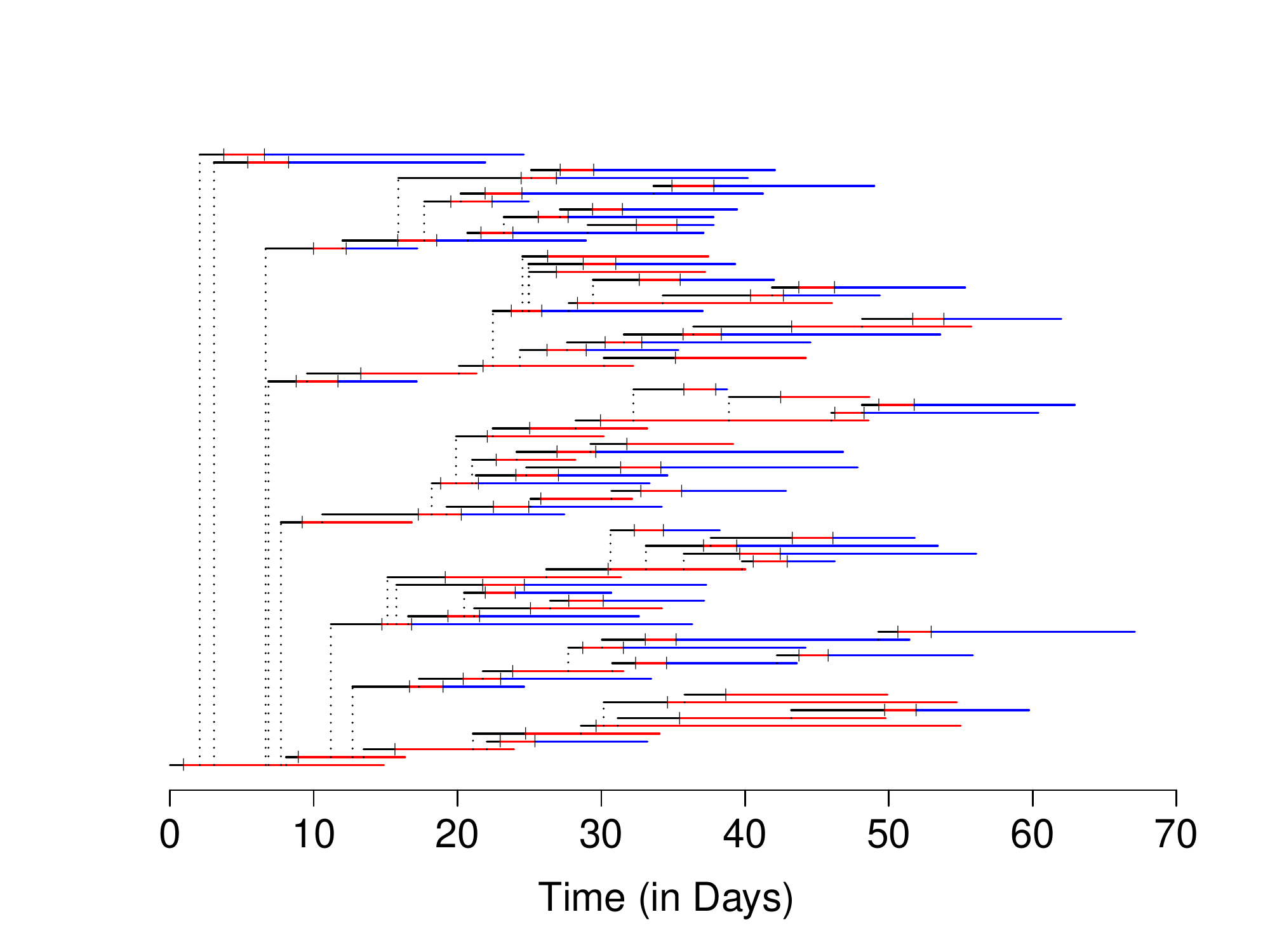}  
\caption{Transmission tree for one simulated epidemic in a population of 100 individuals at baseline parameter values.  Each horizontal line segment represents an infected individual.  The black portion represents the time in the Exposed state, the red portion represents the time in the Infectious (but not Quarantined) state, and the blue portion represents the time in the Quarantine state.  Black vertical dotted lines represent transmission events.}
\label{fig:transtree}
\end{figure}

\end{appendices}

\end{document}